\documentclass[%
 reprint,
showpacs,
 amsmath,amssymb,
 aps,
pre,
floatfix,
]{revtex4-1}

\usepackage{graphicx}
\usepackage{dcolumn}
\usepackage{bm}
\usepackage[caption=false]{subfig}

\begin{document}


\title{Causal Non-Linear Financial Networks}

\author{Pawe\l{} Fiedor}
\email{Pawel.F.Fiedor@ieee.org}
\affiliation{%
 Cracow University of Economics\\
 Rakowicka 27, 31-510 Krak\'{o}w, Poland
}%

\date{\today}

\begin{abstract}
In our previous study we have presented an approach to studying lead--lag effect in financial markets using information and network theories. Methodology presented there, as well as previous studies using Pearson's correlation for the same purpose, approached the concept of lead--lag effect in a naive way. In this paper we further investigate the lead--lag effect in financial markets, this time treating them as causal effects. To incorporate causality in a manner consistent with our previous study, that is including non-linear interdependencies, we base this study on a generalisation of Granger causality in the form of transfer entropy, or equivalently a special case of conditional (partial) mutual information. This way we are able to produce networks of stocks, where directed links represent causal relationships for a specific time lag. We apply this procedure to stocks belonging to the NYSE 100 index for various time lags, to investigate the short-term causality on this market, and to comment on the resulting Bonferroni networks.
\end{abstract}

\maketitle


\section{Introduction}

The study of financial markets as complex adaptive systems has been largely developed by the field of econophysics, as both complexity and adaptiveness of socio-economic systems have been ignored by a majority of mainstream economic studies. Network theory has been particularly useful in analysing financial markets and the interdependencies within them. Such studies started being published around 15 years ago \cite{Mantegna:1999}, and this kind of research is still very much active. Recently the interest of researchers has shifted from the static analysis of the structure of a given market to the analysis of network evolution over time, criticality within networks, or asynchronous relationships within the studied markets. In this paper we analyse the last of the mentioned problems. Standard assumption in economics asserts that time series describing stock returns are unpredictable \cite{Samuelson:1965}. Moreover, the Efficient-Market Hypothesis \cite{Tobin:1969} proposes that all information is reflected in the prices, thus it is not possible to predict future prices based on the past. Within this paradigm there can be no lead--lag effect (where a change in one price leads to a similar (or opposite) change in another price at a specific later time) in financial markets. The Efficient-Market Hypothesis does not hold in practice however \cite{Lo:1988,Shmilovici:2003,Fiedor:2014}. In this case it is plausible that financial data is structured. In this study we try to explore this structure and analyse the New York Stock Exchange, paying attention particularly to causal relationships. Exploring the structure (inefficiencies) of financial markets is of obvious importance to risk analysis for investors and other counterparties.

Studies of financial networks have their roots in tools developed to model physical systems \cite{Mandelbrot:1963,Kadanoff:1971,Mantegna:1991}. Most of past studies have concentrated on correlation structure within financial markets for daily \cite{Mantegna:1999,Cizeau:2001,Forbes:2002,Podobnik:2008,Aste:2010,Kenett:2012meta} and intraday returns \cite{Bonanno:2001,Tumminello:2007,Munnix:2010}. In recent years other measures of similarity have also been used, including Granger causality analysis \cite{Billio:2002}, partial correlation analysis \cite{Kenett:2010}, and information-theoretic concepts \cite{Fiedor:2014a}. All of these methods try to uncover meaningful information in the increasingly complex adaptive systems of financial markets.

As mentioned above, most previous studies analysed synchronous correlations structures of equity returns. Such analyses have shown that financial markets have a nested structure organised in groups defined by economic sector. The correlations can be exchanged for another well-defined similarity measure such as mutual information \cite{Fiedor:2014a} without drastically altering the results. These results are indeed well corroborated, and have been obtained using a variety of methods, including random matrix theory \cite{Laloux:2000}, principal component analysis \cite{Fenn:2011}, hierarchical clustering \cite{Mantegna:1999}, correlation--based networks \cite{Mantegna:1999,Bonanno:2003,Onnela:2003} and mutual information--based networks \cite{Fiedor:2014a}. We are particularly interested in methods used to construct dependency networks. These may be grouped into two distinct categories: threshold--based methods and topological methods. Both are based on the same sample similarity measure. But using the former method a threshold is set on the similarity measure, and a network is constructed in such a way that only links between nodes whose pairwise similarity measure is larger than the threshold value are present in the constructed network. As the threshold value is lowered a more complex hierarchy emerges, where clusters of stocks progressively merge to finally form the whole market. Such networks are very robust with regards to the statistical uncertainty in the similarity measure, but it is difficult to find a single threshold value which accurately displays the nested structure of the studied similarity matrix. Conversely, topological methods construct dependency networks, such as the minimal spanning tree (MST) \cite{Mantegna:1999,Bonanno:2003,Onnela:2003,Fiedor:2014a} or the planar maximally filtered graph (PMFG) \cite{Tumminello:2005,Tumminello:2010,Fiedor:2014a}. These are based on the ranking of empirical similarity measures. The resulting networks are intrinsically hierarchical and therefore easy to present graphically, but such approach is less stable with respect to the statistical uncertainty in the data. Furthermore, such an approach is not sensitive to the statistical significance of the similarity measures \cite{Coronnello:2007}.

Very few inquires have looked into asynchronous financial networks \cite{Huth:2011,Curme:2014} however. The methods above cannot be easily employed to the analysis of directed lagged dependencies in financial markets. The lagged interdependencies within stock returns are generally weak even at short time horizons, therefore an analysis is strongly influenced by the statistical uncertainty of the estimation process, which is not the case for synchronous analysis where most pairwise similarity measures are strongly significant. Thus the use of topological methods is difficult, as these do not take into consideration the values of similarity measure of their statistical significance. There is a danger that many links in such a network created for lagged relationships would be statistically insignificant. On the other hand, threshold methods are difficult to apply as it is difficult to find an appropriate threshold level. Nonetheless threshold methods must be applied, as they are a significantly better than the alternative for this purpose.

In Ref.~\cite{Curme:2014} a method for filtering a lagged correlation matrix into a network of statistically--validated directed links has been introduced. In particular, a $p$-value is associated with each observed lagged--correlation, and a threshold is set on $p$-values (a level of statistical significance corrected for multiple hypothesis testing). In our previous study \cite{Fiedor:2014lag} we extended this analysis to include non-linear relationships. It is well corroborated that financial markets are behaving in a strongly non-linear manner. There is evidence of non-linear dynamics in stock returns \cite{Brock:1991,Qi:1999,McMillan:2001,Sornette:2002,Kim:2002}, market index returns \cite{Franses:1996,Wong:1995,Chen:1996,Wong:1997,Ammermann:2003}, and currency exchange rate changes \cite{Hsieh:1989,Brock:1991,Rose:1991,Brooks:1996,Wu:2003}. Therefore an analysis using Person's correlation, strictly not sensitive to non-linear dependencies, can miss important features of such dynamical systems. We used mutual information ($I$) \cite{Cover:1991} instead of correlation, as it is more general measure, where $I = 0$ if and only if the two studied random variable are strictly independent. Recently we have used mutual information in creating dependency networks of financial markets \cite{Fiedor:2014a}. Additionally, in our previous study we have also shown that a less computationally expensive method for filtering, based on Gamma distribution, may be giving similarly good results as the heterogeneous bootstrap method introduced in Ref.~\cite{Curme:2014}.

Both mentioned studies use a naive way of looking at directed lagged financial networks however. Such analyses, being time-asymmetric, necessarily hint at causal relationships. But an analysis based cross--correlation or mutual information between lagged time series does not deal with causality. It is useful to find such structures of lagged synchronisation, but in this paper we extend this methodology to capture causal relationships in the studied financial markets. This can be done in a natural way on the basis of information theoretic approach presented in our previous study. Mutual information used there can be exchanged for partial mutual information conditioned on the lagged version of the financial instrument, which is representing the effect in the studied pair. Such measure is equivalent to transfer entropy, which is in itself a non-linear generalisation of Granger causality. We can therefore easily extend our analysis to make it more strict in finding causal relationships (in the sense of Granger causality) and not merely lagged structural dependencies. Transfer entropy has been studied extensively in the last decade in a multitude of fields \cite{Chicharro580051}. Transfer entropy has also been used to study financial markets, though mostly in a setting unrelated to networks \cite{Marschinski:2001,Reddy:2009,Dimpfl2012Using,Li:2013}. Additionally most of these studies calculate transfer entropy in a slightly different manner, and in some cases use very peculiar tools and non-standard ways of discretising stock returns. There were some previous studies using transfer entropy to analyse financial networks, but within these either the method of calculating the entropy was different, the validation of links was performed in a different manner or the created networks were constructed differently (e.g. using Minimally Spanning Tree) \cite{Baek:2005,Kwon:2008,Junior:2013}. Thus none of these have used the precise methodology or had the same aim as we present in this study.

The paper is organised as follows. In Section~2 we present a method used to filter and validate statistically significant causal connections based on the partial mutual information (transfer entropy) between financial instruments with appropriate time lags. In Section~3 we analyse the structure of NYSE 100 at various time lags using the presented methodology. In Sect.~4 we discuss the results. In Sect.~5 we conclude the study and propose further research.

\section{Methods}

Here we present a methodology for statistically validating lagged partial mutual information (transfer entropy) for the purpose of network analysis of causality within financial markets. We begin the analysis by preparing a matrix of logarithmic returns over given intraday time--horizons. Let us denote the most recent price for stock $n$ occurring on time $t$ during the studied period by $p_n(t)$. Additionally, $\tau$ is the time horizon. Then for each stock the logarithmic returns are sampled,
\begin{equation}
 r_{n,t} = \log(p_n(t)) - \log(p_n(t-\tau)),
\end{equation}
every $\tau$ minutes (seconds) throughout the studied period. These time series build columns of matrix $R$. $R$ is consequently filtered into two matrices, $A$ and $B$, in which returns during the last period of length $\lambda$ are excluded from $A$ and returns during the first period of length $\lambda$ are excluded from $B$ (thus $\lambda$ denotes the lag). From these matrices an empirical transfer entropy matrix $C$ is constructed using the partial mutual information of columns from $A$ and $B$,
\begin{equation}
C_{m,n} = I(A_{m},B_{n}|A_{n}),
\label{eqn:corr_matrix}
\end{equation}
where $I(X,Y)$ is the mutual information between $X$ and $Y$, $I(X,Y|Z)$ is the conditional (partial) mutual information \cite{Frenzel:2007} between $X$ and $Y$ conditioned on $Z$, and $A_{m}$ denotes column $m$ of $A$. Note that $C_{m,m}=0$.

Here we show that the above is measuring generalised Granger causality. Transfer entropy is a non-parametric statistic measuring the amount information (in Shannon's sense) transferred between two random processes (transfer entropy is directional). Transfer entropy is thus a measure of Granger causality, but a more general one, sensitive to non-linear interactions \cite{Schindler:2007}. Assuming the data as presented above we can define transfer entropy as:
\begin{equation}
T_{m\to{}n}=H(B_{n}|A_{n})-H(B_{n}|A_{n},A_{m}),
\end{equation}
where $H(X)$ is Shannon's entropy and $H(X|Y)$ denotes conditional Shannon's entropy. Transfer entropy is equivalent to a specific conditional (partial) mutual information \cite{Schindler:2007}:
\begin{equation}
T_{m\to{}n}=I(A_{m},B_{n}|A_{n}).
\end{equation}

To apply the above in practice we need an estimator of entropy (mutual information can be defined in terms of entropy). We note that for easy estimation we need discrete data, and while stock returns are discrete, their resolution is much too high for practical purposes, thus we need to discretise them. For discussion of this step see below and Refs.~\cite{Fiedor:2014,Fiedor:2014a}. There is a large number of estimators of entropy, for details please see Refs.~\cite{Beirlant:1997,Darbellay:1999,Paninski:2003,Daub:2004,Nemenman:2004}. In this study we continue to use the Schurmann--Grassberger estimate of entropy, which we have applied in our previous study \cite{Bonachela:2008}. The Schurmann--Grassberger estimator is a Bayesian parametric procedure, which assumes samples distributed according to Dirichlet distribution:
\begin{equation}
\begin{split}
&\hat{H}(X)=\frac{1}{m+|\chi|N} \\
&\sum_{x\in\chi}{(\#(x)+N)(\psi{}(m+|\chi|N+1)-\psi(\#(x)+N+1))},
\end{split}
\end{equation}
where $\#(x)$ is the number of data points with value $x$, $|\chi|$ is the number of bins from the discretisation step, $m$ is the sample size, and $\psi(z)=\mathrm{d}\ln{\Gamma(z)}/\mathrm{d}z$ is the digamma function. The Schurmann--Grassberger estimator assumes $N=1/|\chi|$ as the prior \cite{Schurmann:1996}.

The matrix $C$ can be seen as a weighted adjacency matrix for a fully connected, directed graph. As stated above, such matrix needs to be filtered. To find a threshold of statistical significance Curme et al. \cite{Curme:2014} apply a shuffling technique \cite{Efron:1993}. The rows of $A$ are shuffled repeatedly without replacement in order to create a large number of surrogate time series. These are then validated by $p$-value adjusted to account for multiple comparisons. Curme et al. \cite{Curme:2014} use the conservative Bonferroni correction ($p/N^2$). For $N=100$ and the unadjusted $p$-value equal to $0.01$ it gives $0.01/100^2$, which requires the construction of $10^6$ independently shuffled surrogate time series. The same can be done for methodology based on mutual information and transfer entropy. But we find the computational requirements of this procedure to be prohibitively large for studies of large networks (at least for practical, real-time applications), and financial markets usually consist of hundreds of stocks. Nonetheless, a less computationally expensive method has been presented in our previous study, without introducing very strong assumptions (here we note that the bootstrap method is, at least in principle, better, if computation time is not an issue, as it takes into account the heterogeneousness of the studied time series). It has been shown that mutual information between independent random variables ($X$ \& $Y$), when estimated from relative frequencies, follows a very good approximation of Gamma distribution with parameters $a=(D)/2$ and $b=1/(N\ln{}2)$ \cite{Goebel:2005,Dawy:2006}:
\begin{equation}
I(X,Y)\sim\Gamma(\frac{D}{2},\frac{1}{N\ln{}2}),
\end{equation}
where $N$ is the sample size and $D$ denote the number of degrees of freedom (dependent on the alphabet used for the studied discrete time series). This has been explained in Ref.~\cite{Fiedor:2014lag}.

Therefore, to determine the significance of $I(A_m,B_n)$ from a sample study of length $N$ at a significance level $p$, we check the condition:
\begin{equation}
I(A_m,B_n)\geq\Gamma_{1-p}(\frac{D}{2},\frac{1}{N\ln{}2}),
\end{equation}
where $\Gamma_{1-p}(a,b)$ denotes the $(1-p)$-quantile of the Gamma distribution. The same holds for conditional mutual information, or transfer entropy, but obviously the quantiles themselves will have different values as the number of degrees of freedom is higher in transfer entropy as it is in mutual information. We note that besides bootstrapping and validating based on Gamma distribution, a threshold can also be set by the analysts based on their experience and practical needs. This being a less formal approach is not considered here. Here we also note that most of the results presented in this study would remain virtually unchanged (with the exception of the actual number of significant links) by a choice of a slightly different threshold value, thus this step is of relatively mild importance to most applications (where analysts would concentrate on the most significant links and not the ones slightly above threshold).

\section{Empirical Application}

To compare the results of this study with our previous study investigating lead--lag effect \cite{Fiedor:2014lag} we use the same data as in that paper. We have taken log returns for 98 securities out of 100 which constitute the NYSE 100, excluding two with incomplete data. These log returns are intraday (one-minute intervals) and cover 15 days between the 21st October 2013 and the 8th November 2013. The data has been downloaded from Google Finance database available at http://www.google.com/finance/, and has been transformed in the standard way for analysing price movements, so that the data points are the log ratios between consecutive closing prices for a given period, as defined above, and those data points are, for the purpose of estimating mutual information, discretised into four distinct states. The states represent quartiles. This design means that the model has no unnecessary parameters and has proved to be very efficient \cite{Steuer:2001,Navet:2008,Fiedor:2014}. The choice of quartiles over other quantiles is largely irrelevant, see the discussion in Ref.~\cite{Fiedor:2014}.

We have set the $p$-value to $0.01$ and corrected it using conservative Bonferroni correction obtaining the corrected $p$-value ($\alpha$) of roughly $\alpha=10^{-6}$. Bonferroni correction is thought to give conservative results. We use the appropriate Gamma distribution for the validation of transfer entropy. We investigate varying time lags $\lambda$ between one minute and one hour, but keep the $\tau$ as one minute throughout the study as not to unnecessarily lose the fidelity in the data.

\begin{figure}[tbh]
\centering
\includegraphics[width=0.4\textwidth]{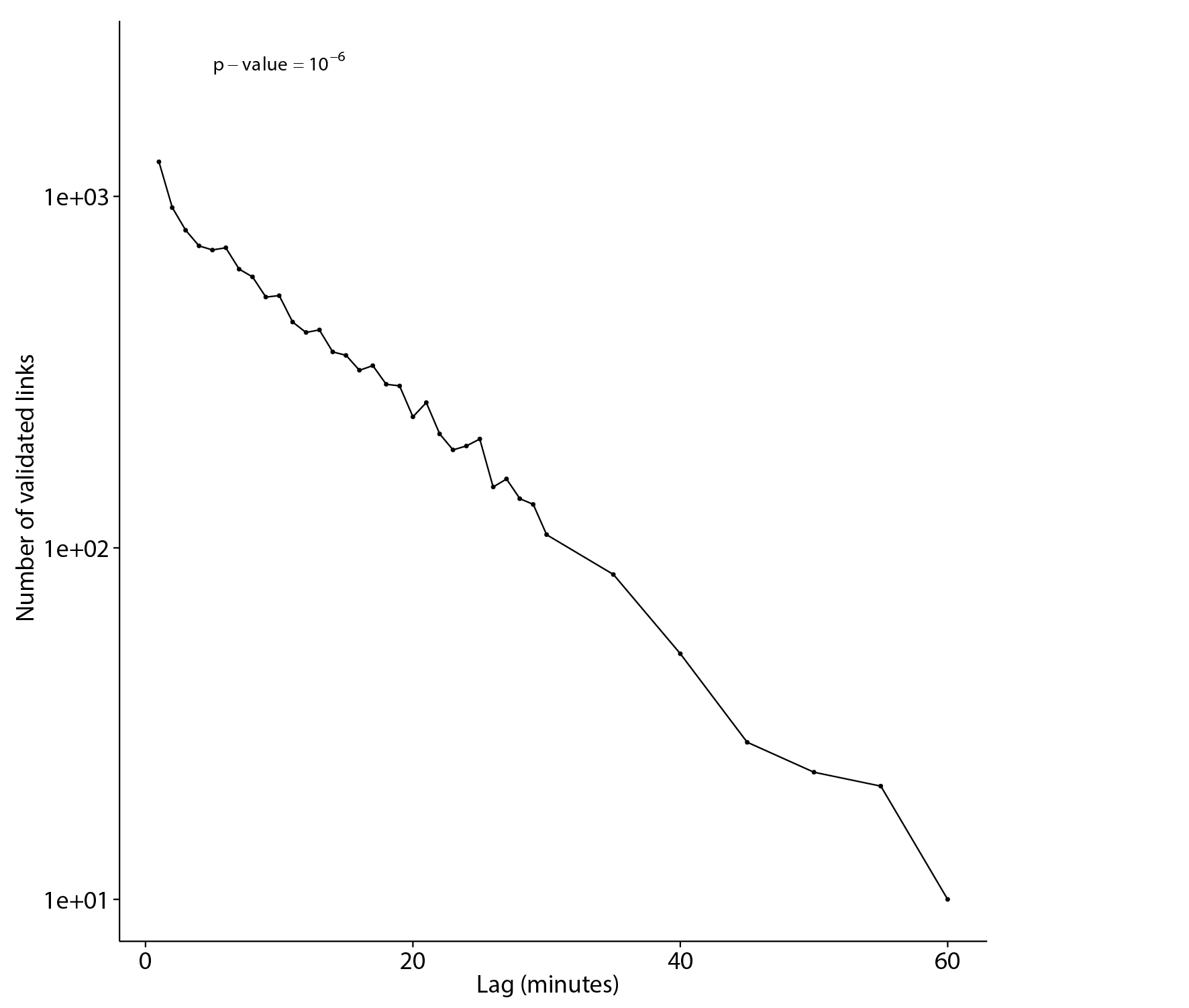}
\caption{Number of pairs of NYSE 100 stocks (out of 9506) with statistically significant transfer entropy--based links for different values of time lag $\lambda$ (with price sampling frequency $\tau$ equal to one minute). Transfer entropy is validated at the specified adjusted $p$-value of $10^{-6}$ for multiple comparisons. We observe a slow decay with increasing time lag up to one hour.}
\label{fig:validated}
\end{figure}

First, we show the relationship between the number of statistically significant lead--lag relationships in the studied market and the time lag $\lambda$. We expect to find a decay in the number of statistically significant lead--lag relationships with growing time lag. In Fig.~\ref{fig:validated} we present the number of validated (statistically significant) transfer entropy--based links for a given time lag of $\lambda$, with price sampling frequency $\tau$ equal to one minute, for intraday stock returns, based on the presented methodology. The links are validated by Gamma distribution at adjusted $p$-value of $10^{-6}$, thus we create Bonferroni networks, similarly to Refs.~\cite{Curme:2014,Fiedor:2014lag}.

\begin{figure}[tbh]
\centering
\includegraphics[width=0.4\textwidth]{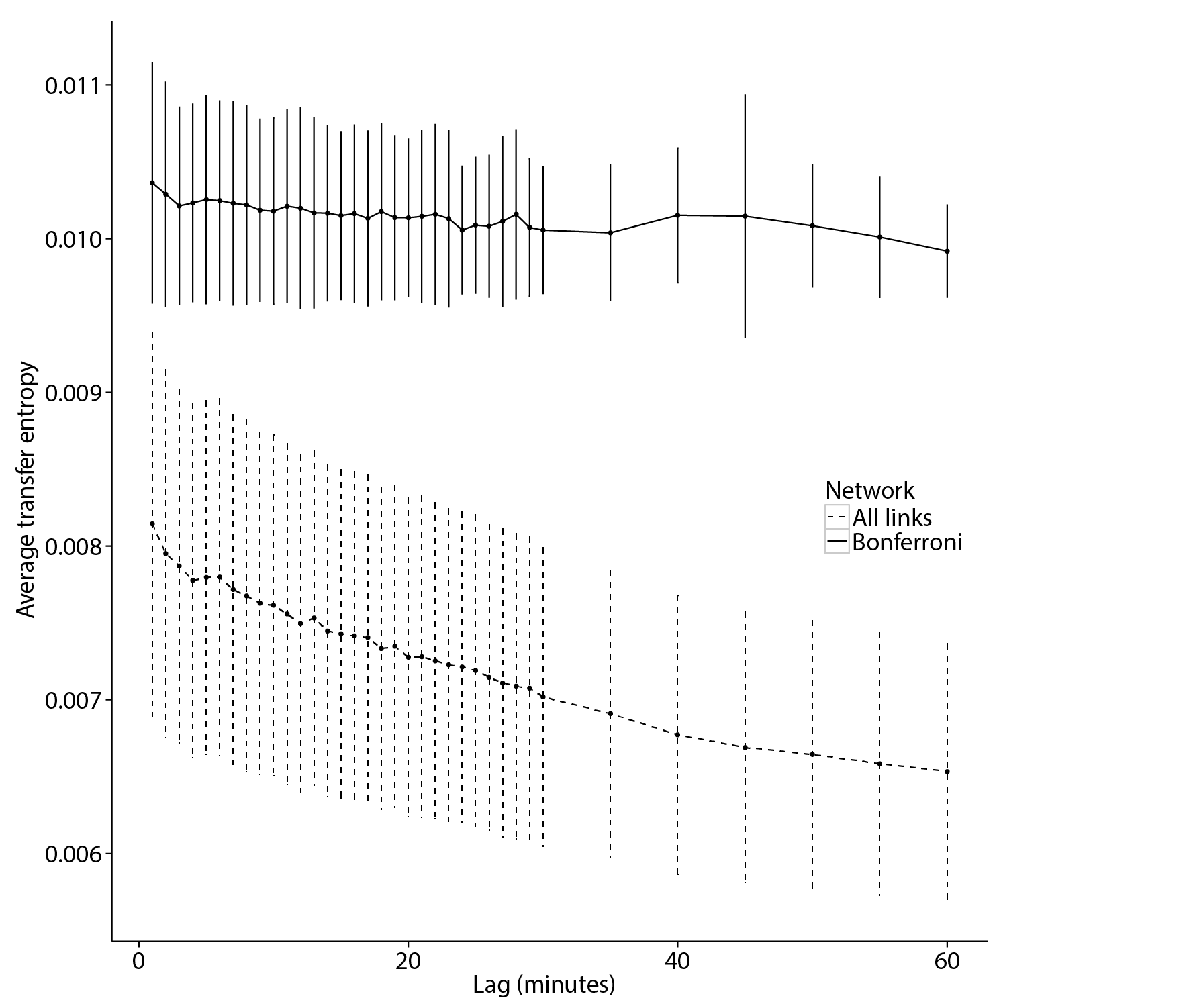}
\caption{Average magnitude of transfer entropy in Bonferroni and full unfiltered networks. Magnitudes appear to be steady for Bonferroni networks and decline for unfiltered networks. Error bars represent plus-or-minus one standard deviation. Interestingly, studies based on correlation have shown these to be steady for unfiltered networks.}
\label{fig:magnitude}
\end{figure}

\begin{figure*}%
\centering
\subfloat[][]{%
\label{fig:hist_dist-a}%
\includegraphics[width=0.5\textwidth]{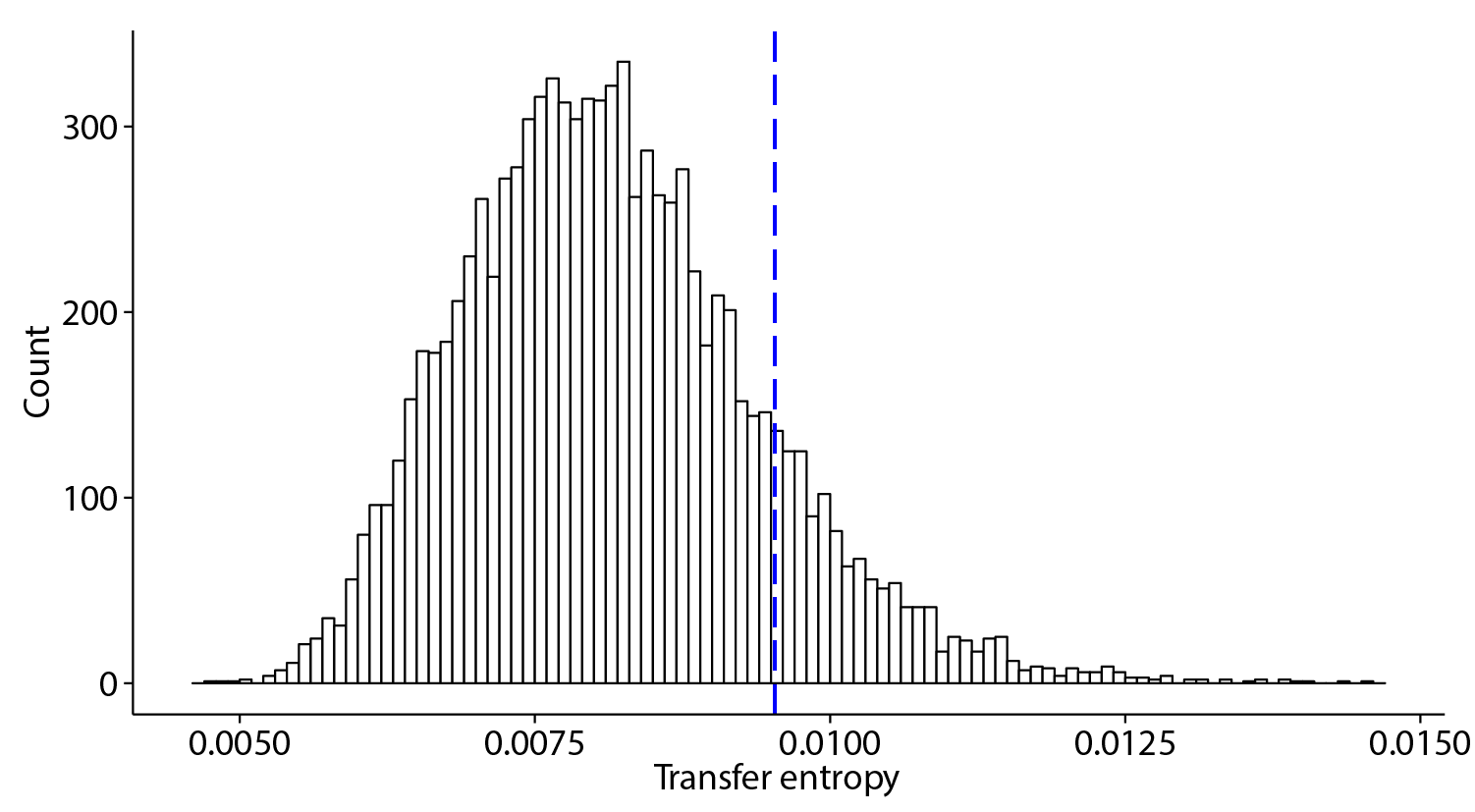}}%
\subfloat[][]{%
\label{fig:hist_dist-b}%
\includegraphics[width=0.5\textwidth]{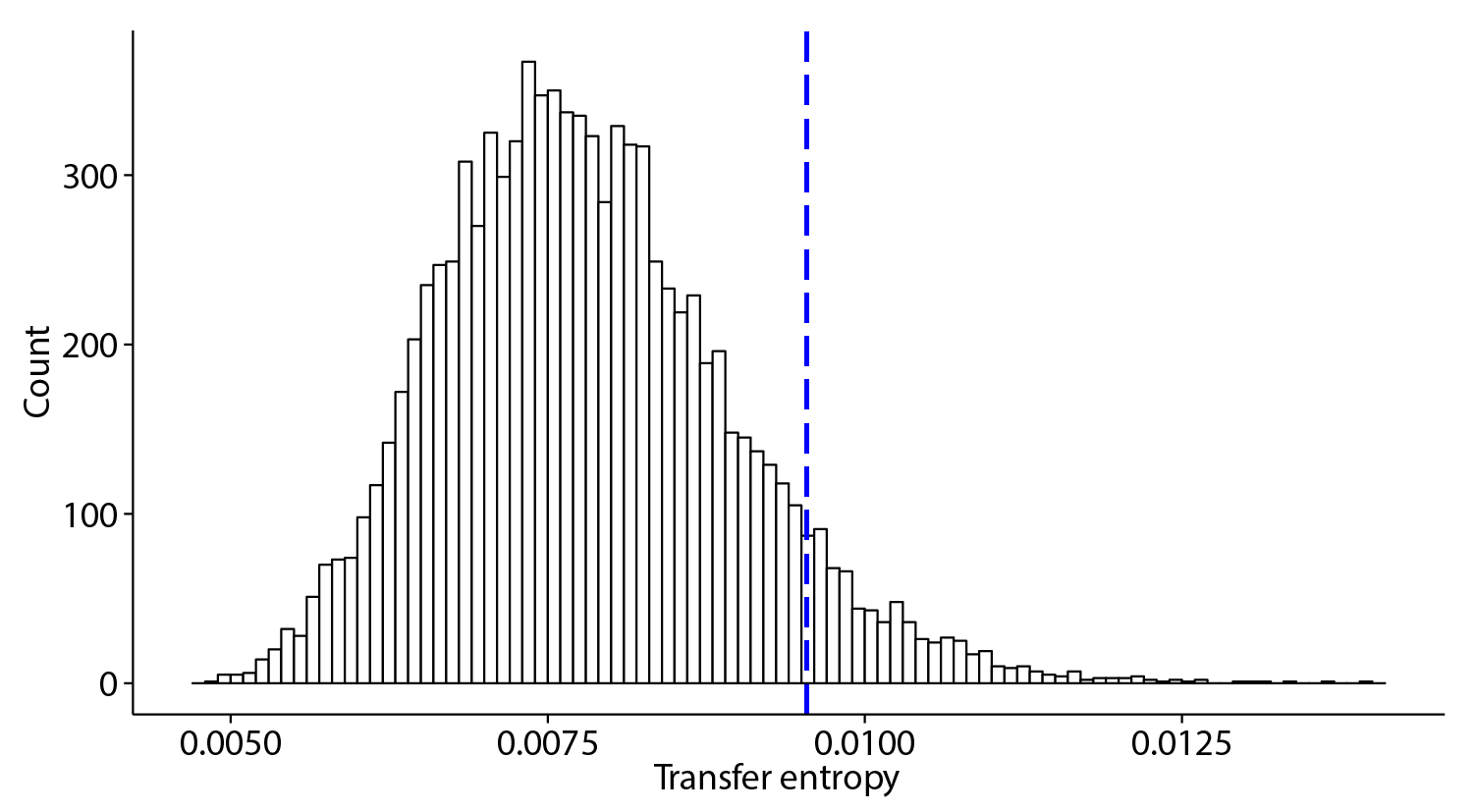}}\\
\subfloat[][]{%
\label{fig:hist_dist-c}%
\includegraphics[width=0.5\textwidth]{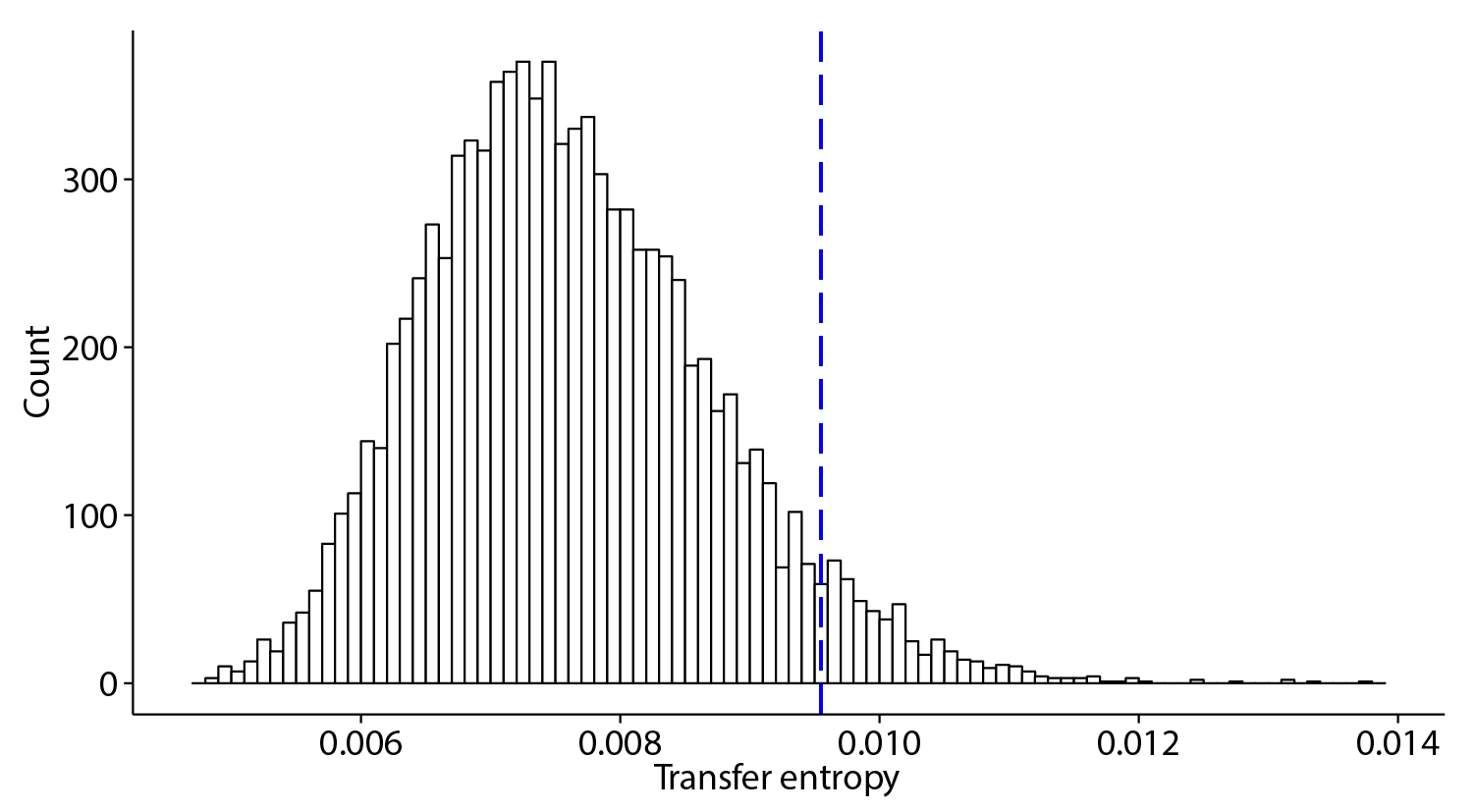}}%
\subfloat[][]{%
\label{fig:hist_dist-d}%
\includegraphics[width=0.5\textwidth]{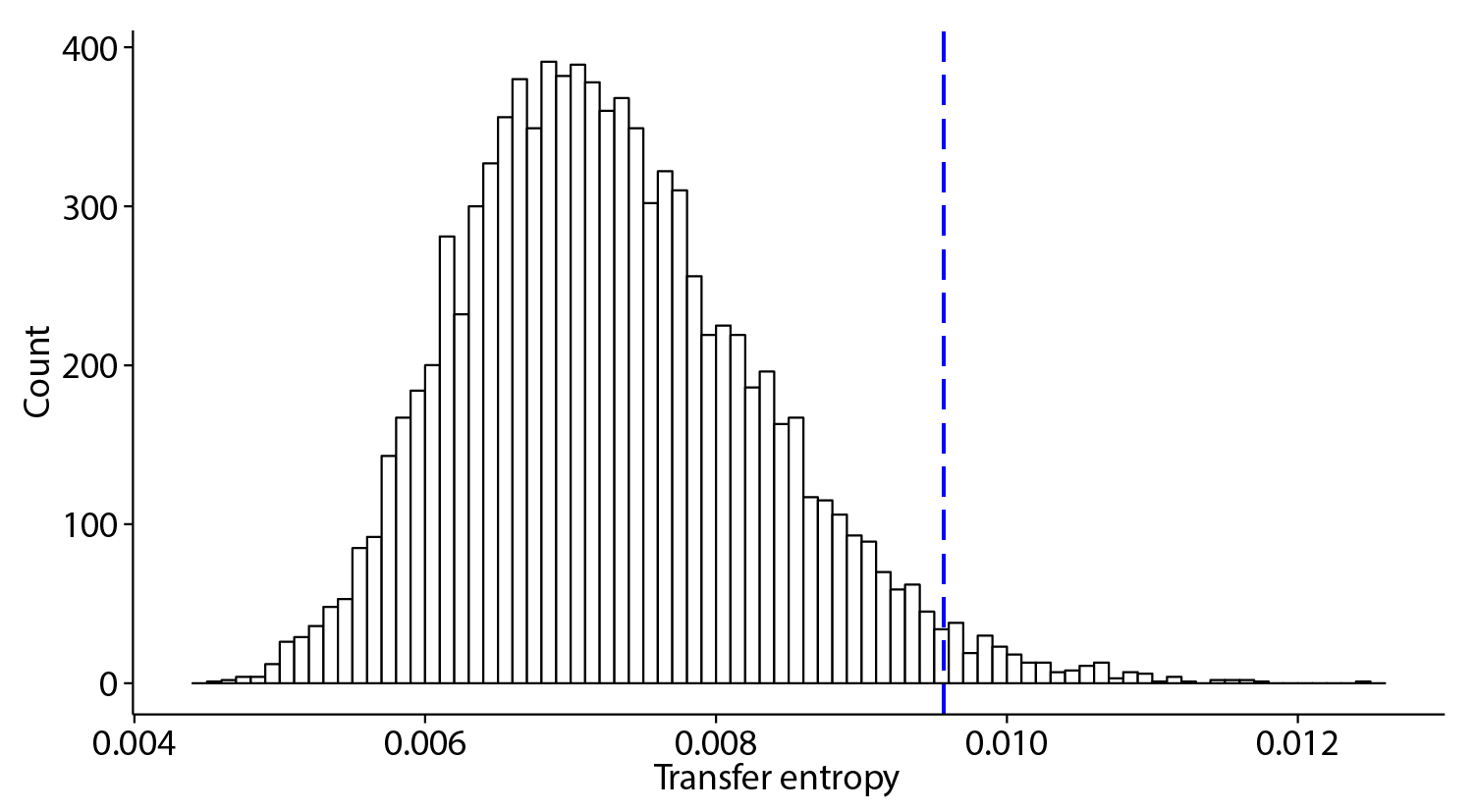}}\\
\subfloat[][]{%
\label{fig:hist_dist-e}%
\includegraphics[width=0.5\textwidth]{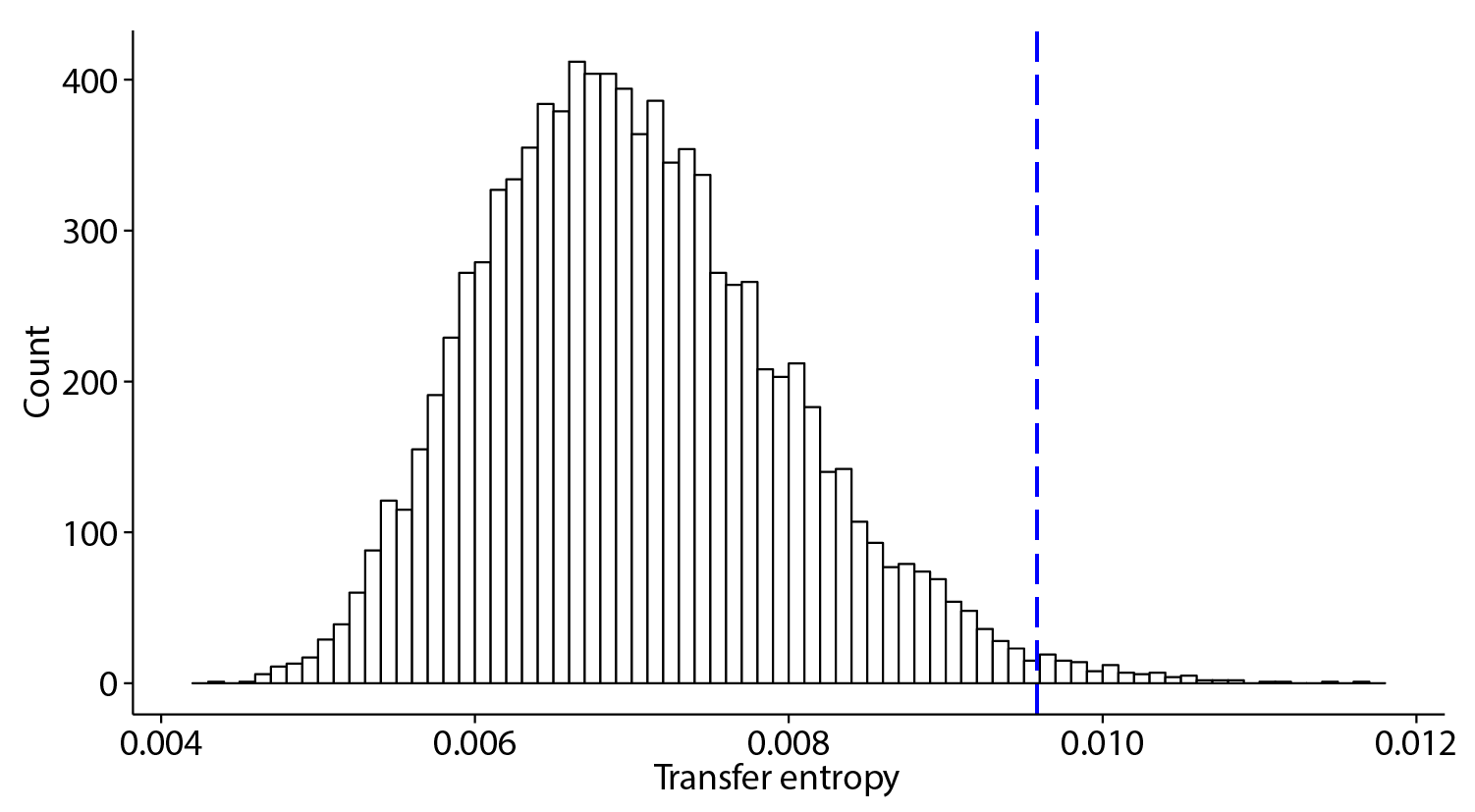}}%
\subfloat[][]{%
\label{fig:hist_dist-f}%
\includegraphics[width=0.5\textwidth]{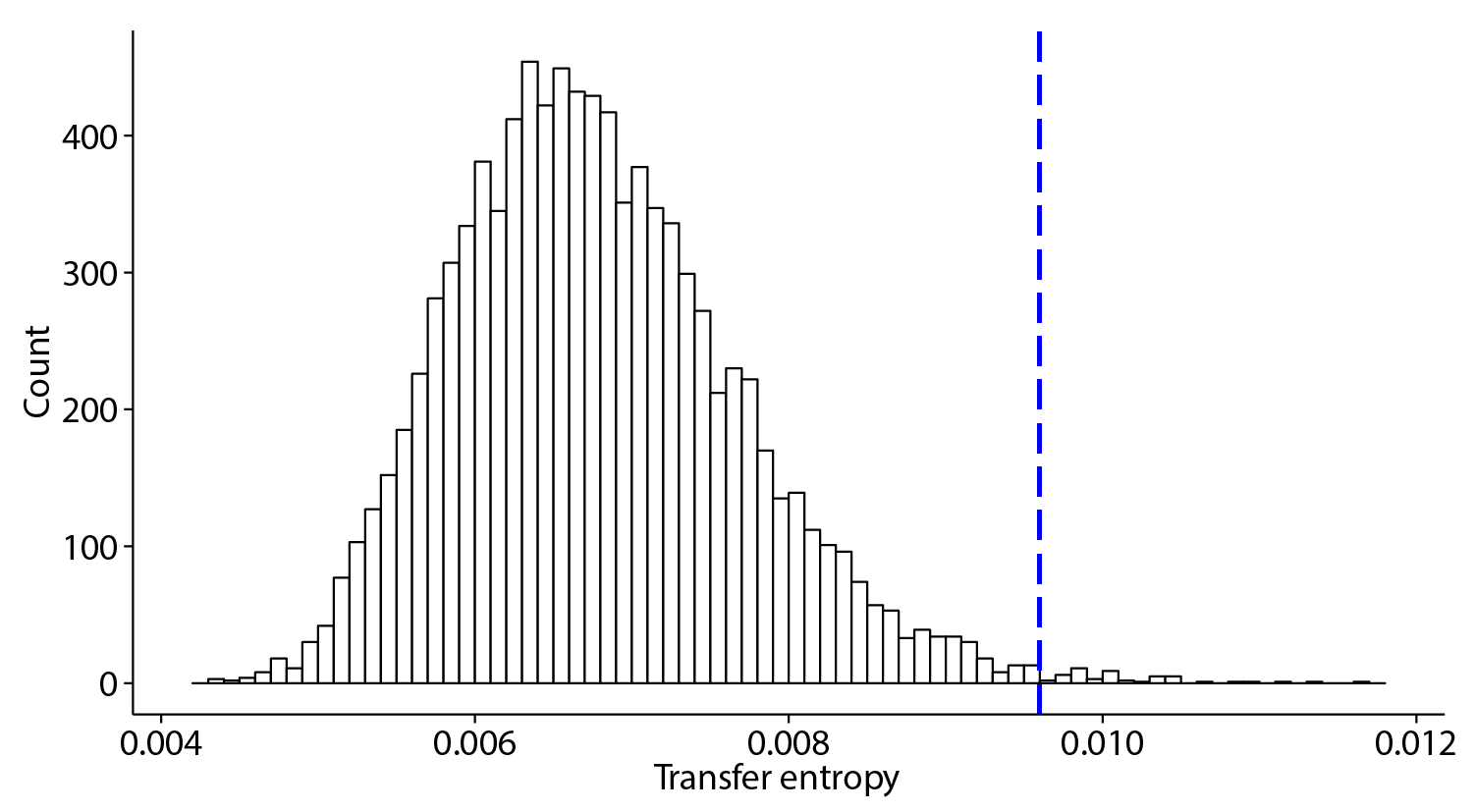}}%
\caption[Degree]{Transfer entropy distribution (with statistical significance threshold shown) for all pairs of NYSE 100 stocks for time lag $\lambda$ of:
\subref{fig:hist_dist-a} one minute;
\subref{fig:hist_dist-b} five minutes;
\subref{fig:hist_dist-c} ten minutes;
\subref{fig:hist_dist-d} twenty minutes;
\subref{fig:hist_dist-e} thirty minutes; and
\subref{fig:hist_dist-f} forty minutes. These distributions appears to be approximated by skewed Normal distribution, and are shifting to the left with increasing time lag $\lambda$.}%
\label{fig:tent}%
\end{figure*}

To see the effect of the validation as shown in Fig.~\ref{fig:magnitude} we present the average transfer entropy for all pairs in the studied set (9506) and for the Bonferroni networks, for varying time lag $\lambda$. The error bars present plus-or-minus one standard deviation. Further, in Fig.~\ref{fig:tent} we present histograms representing distributions of transfer entropy between all pairs of stocks in the studied set, together with the threshold value above which the transfer entropy is statistically significant. These are shown for time lags $\lambda$ of one, five, ten, twenty, thirty and forty minutes.

\begin{figure}[tbh]
\centering
\includegraphics[width=0.4\textwidth]{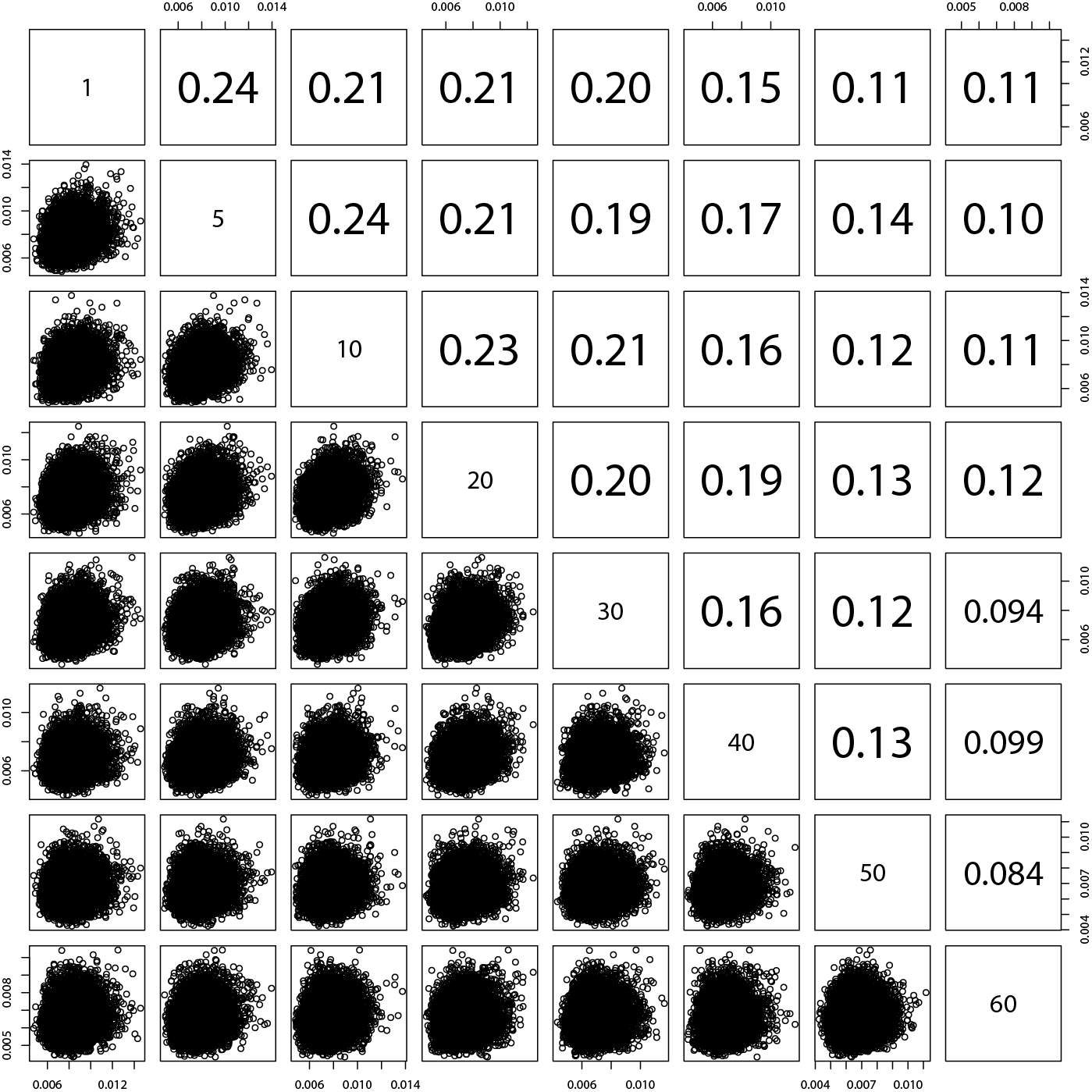}
\caption{Pearson's correlation coefficients between transfer entropy for all pairs of studied stocks for time lag $\lambda$ of one, five, ten, twenty, thirty, forty, fifty, and sixty minutes. There is only a slight positive correlation, hinting that the relationship between stocks at one time lag will not necessarily guarantee that the same relationship at different time lag will be statistically significant.}
\label{fig:correlation}
\end{figure}

In Fig.~\ref{fig:correlation} we present Pearson's correlation coefficients between transfer entropy for all pairs of studied stocks for time lag $\lambda$ of one, five, ten, twenty, thirty, forty, fifty, and sixty minutes. In this way we are able to see whether the relationships at small time lags persist at different, higher, time lags, or whether these are largely unrelated.

\begin{figure*}%
\centering
\subfloat[][]{%
\label{fig:te_netw-a}%
\includegraphics[width=0.3\textwidth]{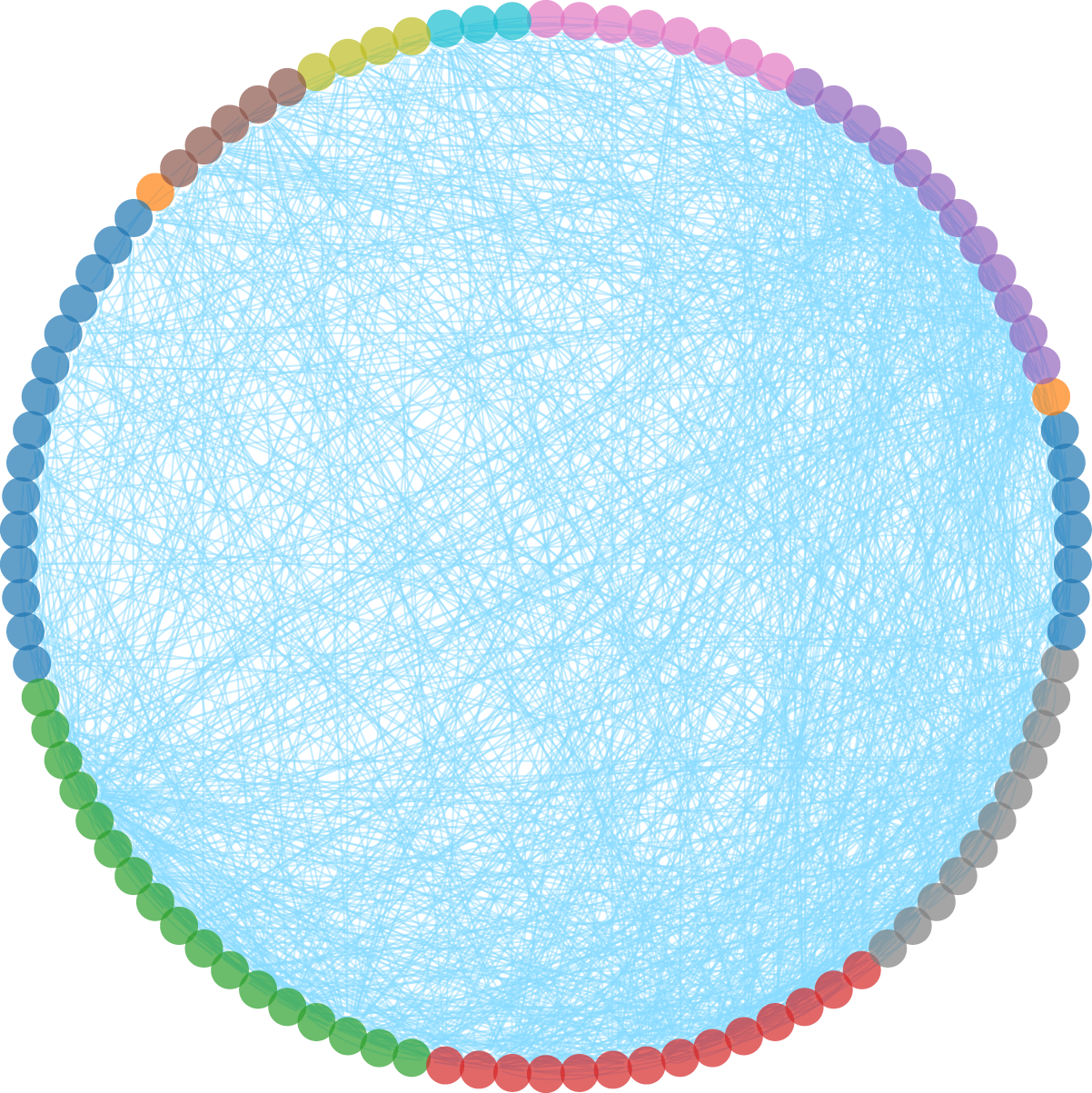}}%
\subfloat[][]{%
\label{fig:te_netw-b}%
\includegraphics[width=0.3\textwidth]{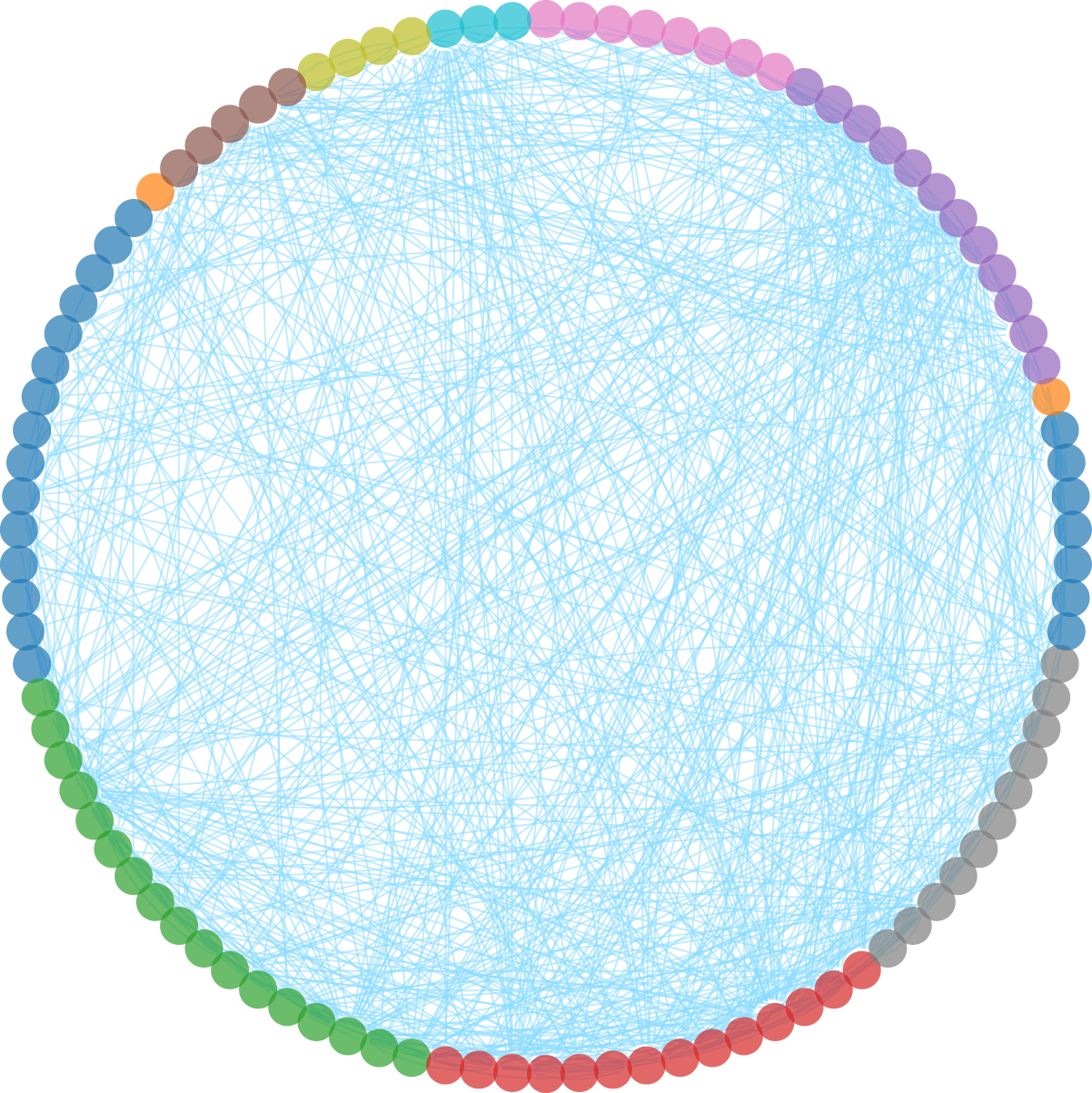}}%
\subfloat[][]{%
\label{fig:te_netw-c}%
\includegraphics[width=0.3\textwidth]{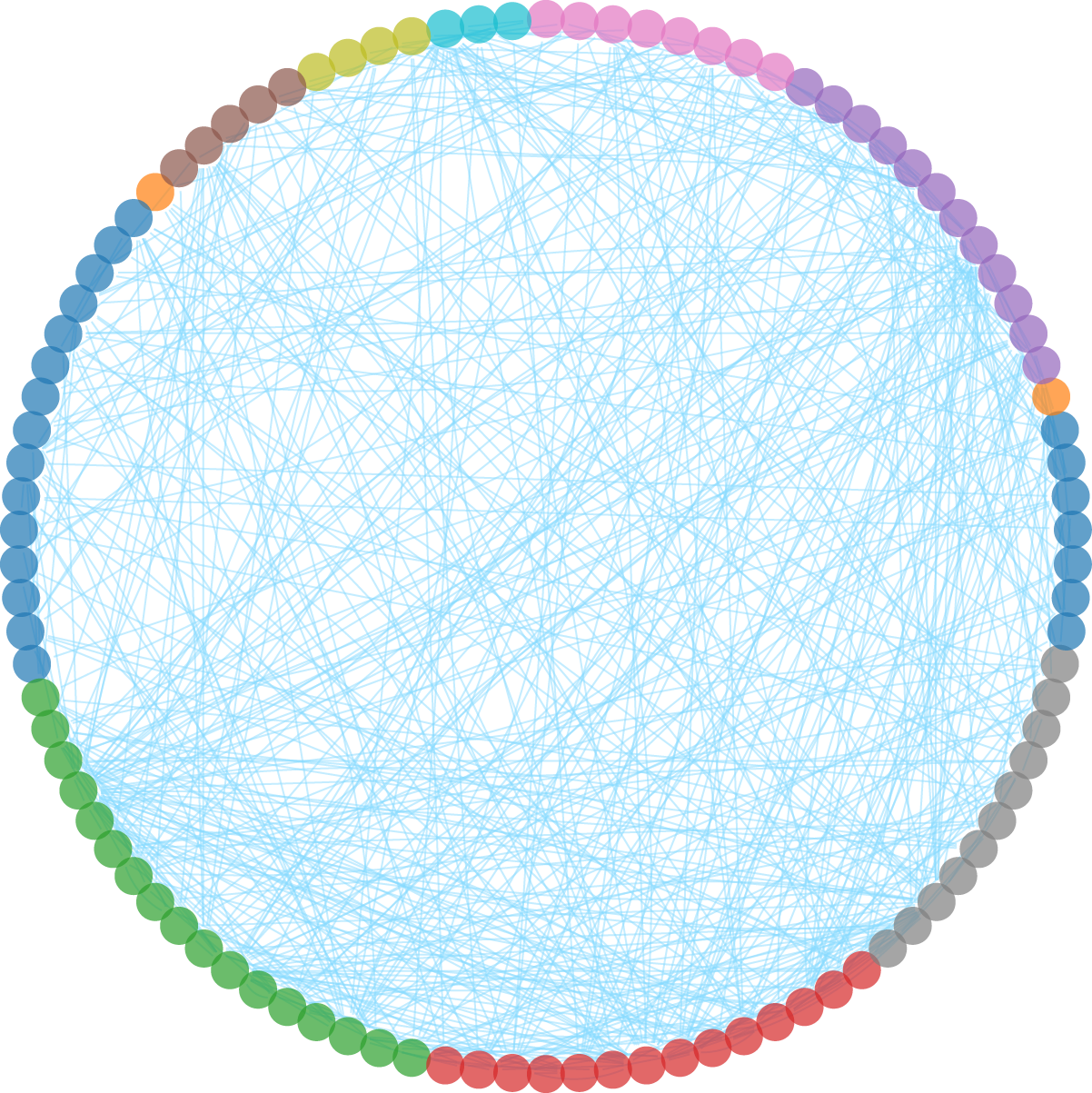}}\\
\subfloat[][]{%
\label{fig:te_netw-d}%
\includegraphics[width=0.3\textwidth]{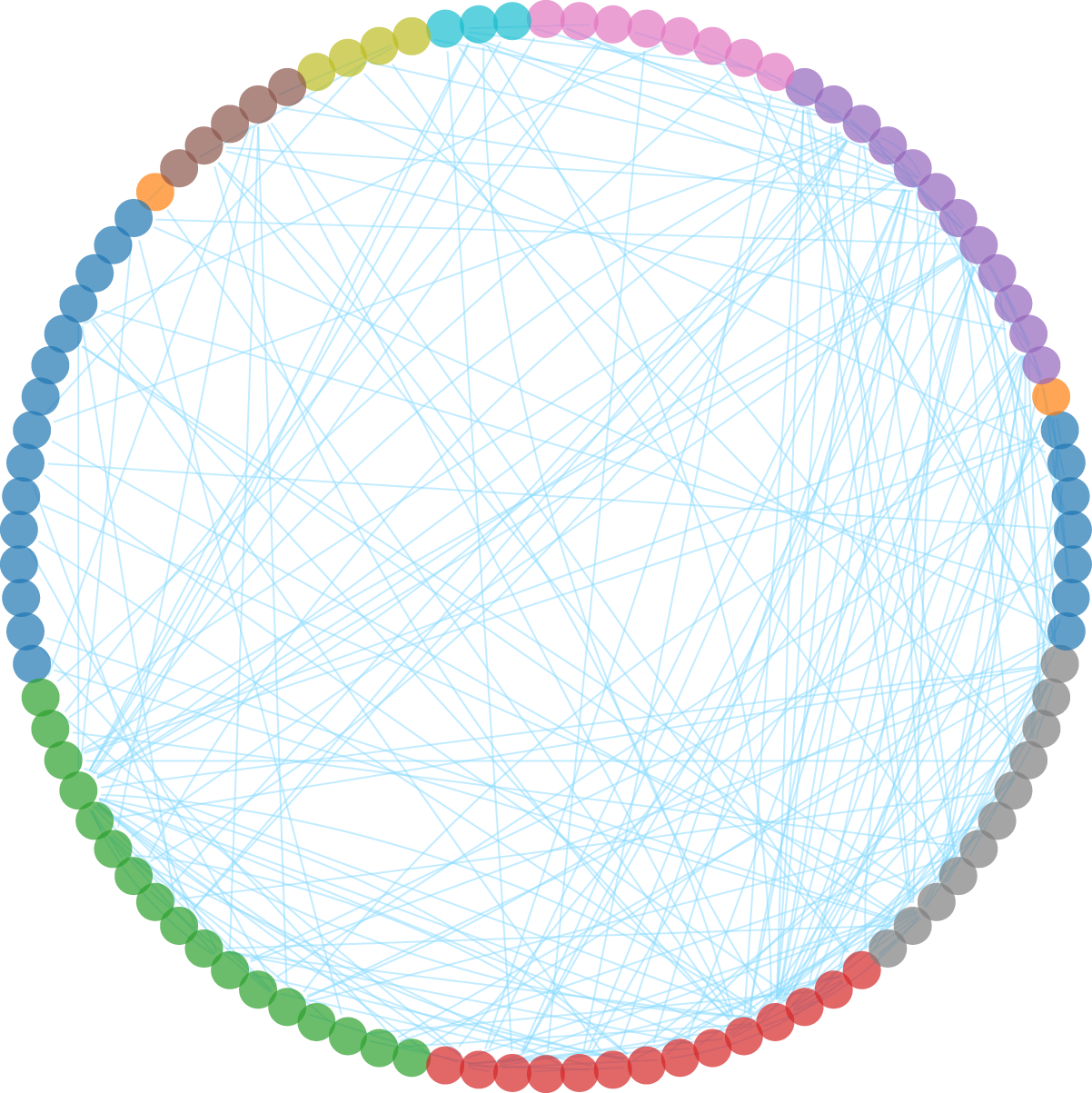}}%
\subfloat[][]{%
\label{fig:te_netw-e}%
\includegraphics[width=0.3\textwidth]{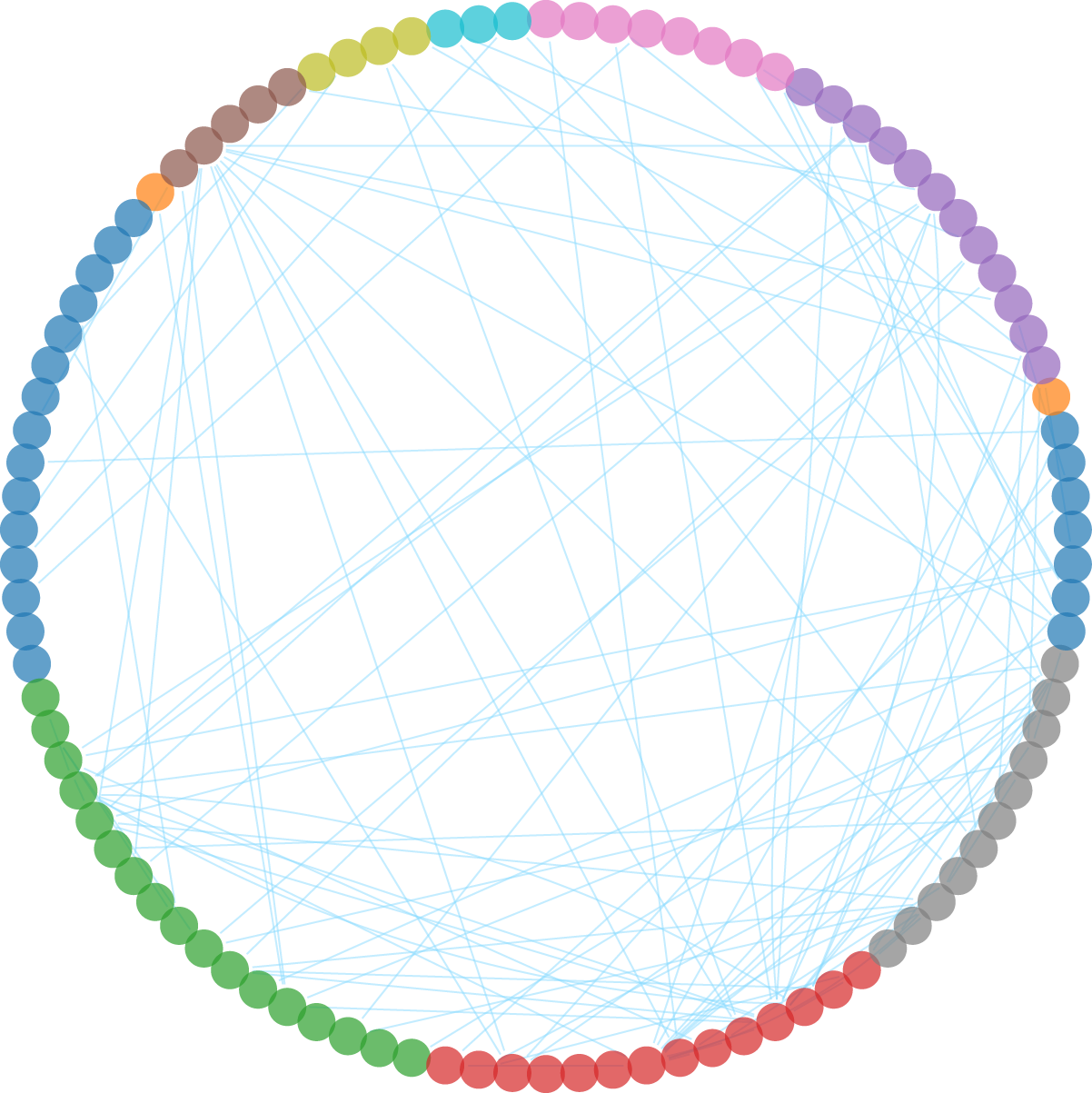}}%
\subfloat[][]{%
\label{fig:te_netw-f}%
\includegraphics[width=0.3\textwidth]{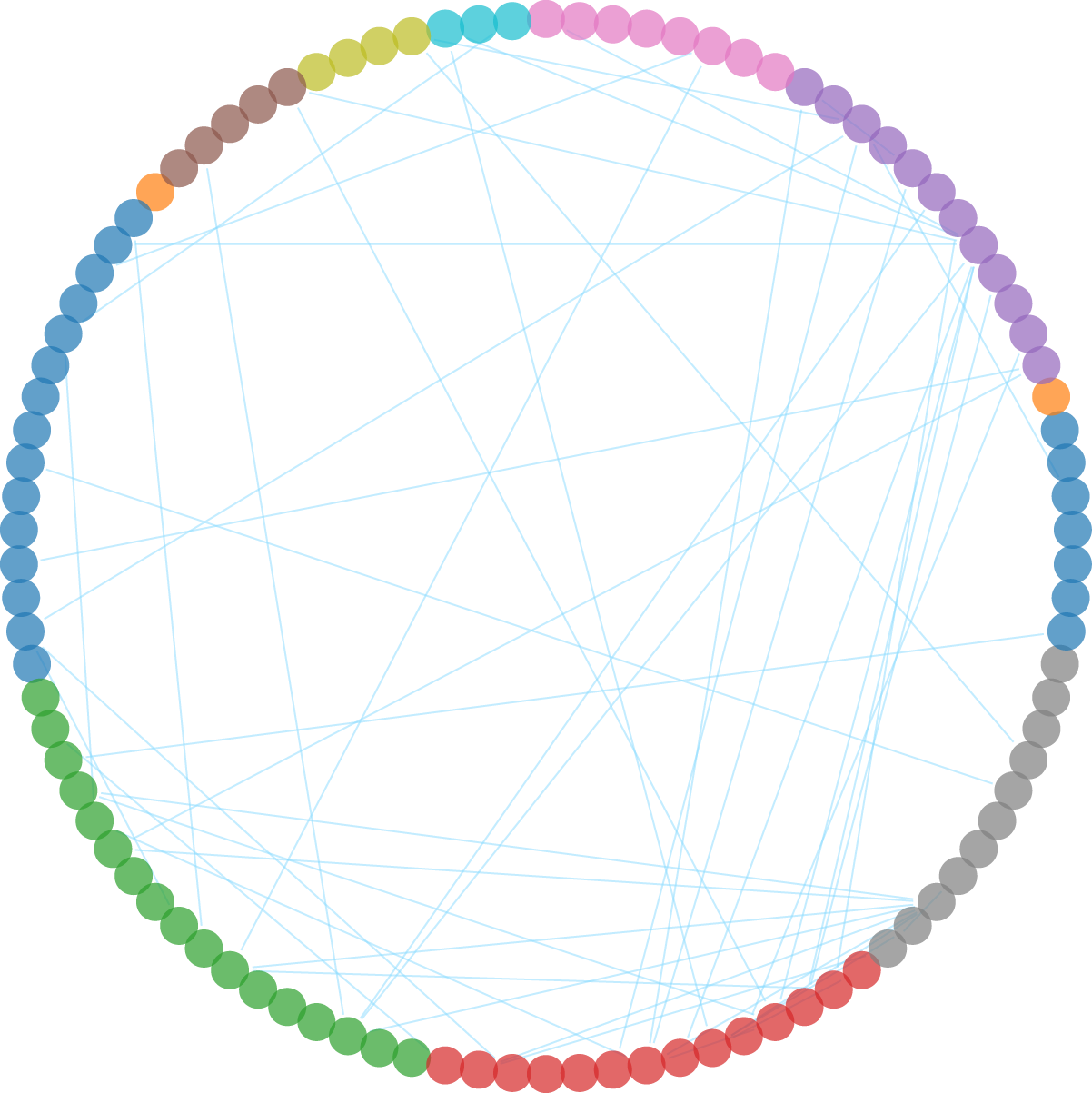}}%
\caption[Degree]{Bonferroni networks of statistically significant transfer entropy between pairs of NYSE 100 stocks for time lag $\lambda$ of:
\subref{fig:te_netw-a} one minute;
\subref{fig:te_netw-b} five minutes;
\subref{fig:te_netw-c} ten minutes;
\subref{fig:te_netw-d} twenty minutes;
\subref{fig:te_netw-e} thirty minutes; and
\subref{fig:te_netw-f} forty minutes. The decay in statistically significant relationships with increasing $\lambda$ is clearly visible.}%
\label{fig:te_netw}%
\end{figure*}

Finally, in Fig.~\ref{fig:te_netw} we present the Bonferroni networks themselves, that is networks consisting only of statistically significant transfer entropy relationships. The stocks are grouped by economic sector as defined by the New York Stock Exchange (the 98 stocks are divided into 12 sectors). The position of all stocks is the same in all presented networks. The networks are shown for time lag $\lambda$ of one, five, ten, twenty, thirty and forty minutes.

\begin{figure*}%
\centering
\subfloat[][]{%
\label{fig:bonf_dd-a}%
\includegraphics[width=0.5\textwidth]{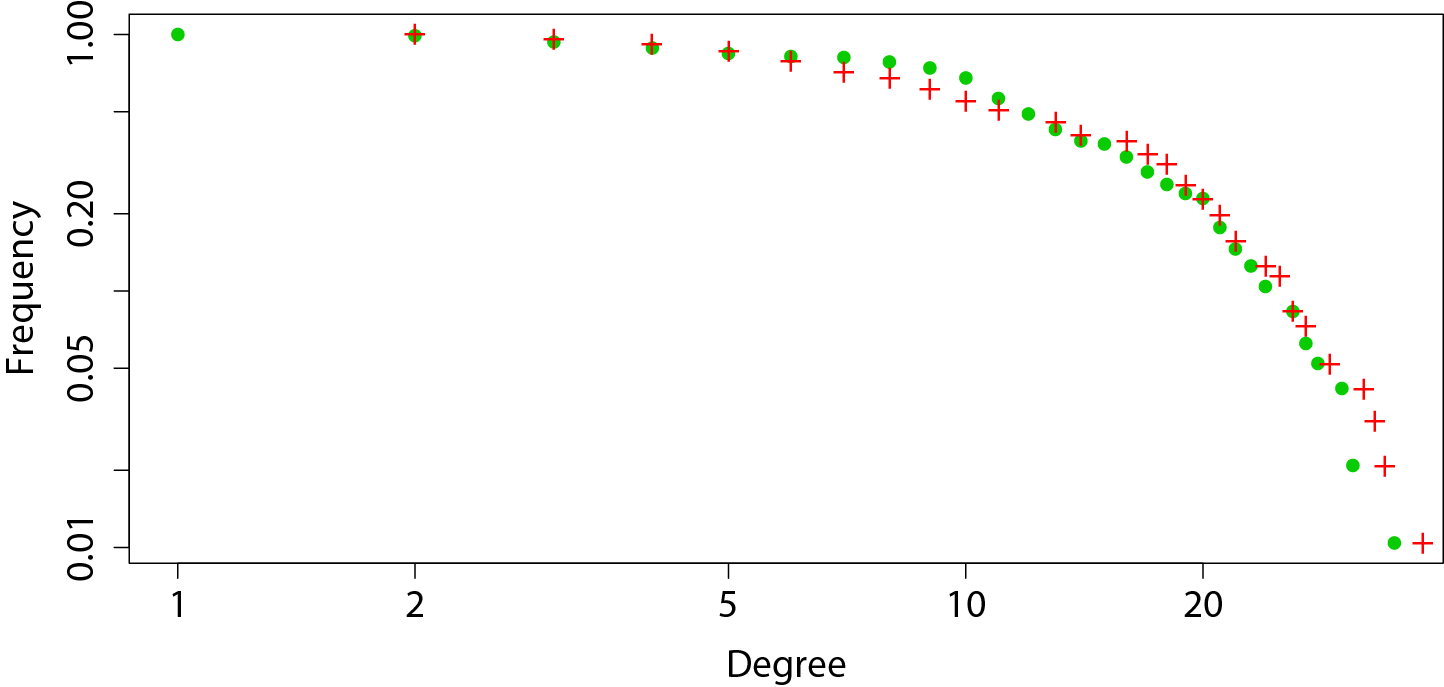}}%
\subfloat[][]{%
\label{fig:bonf_dd-b}%
\includegraphics[width=0.5\textwidth]{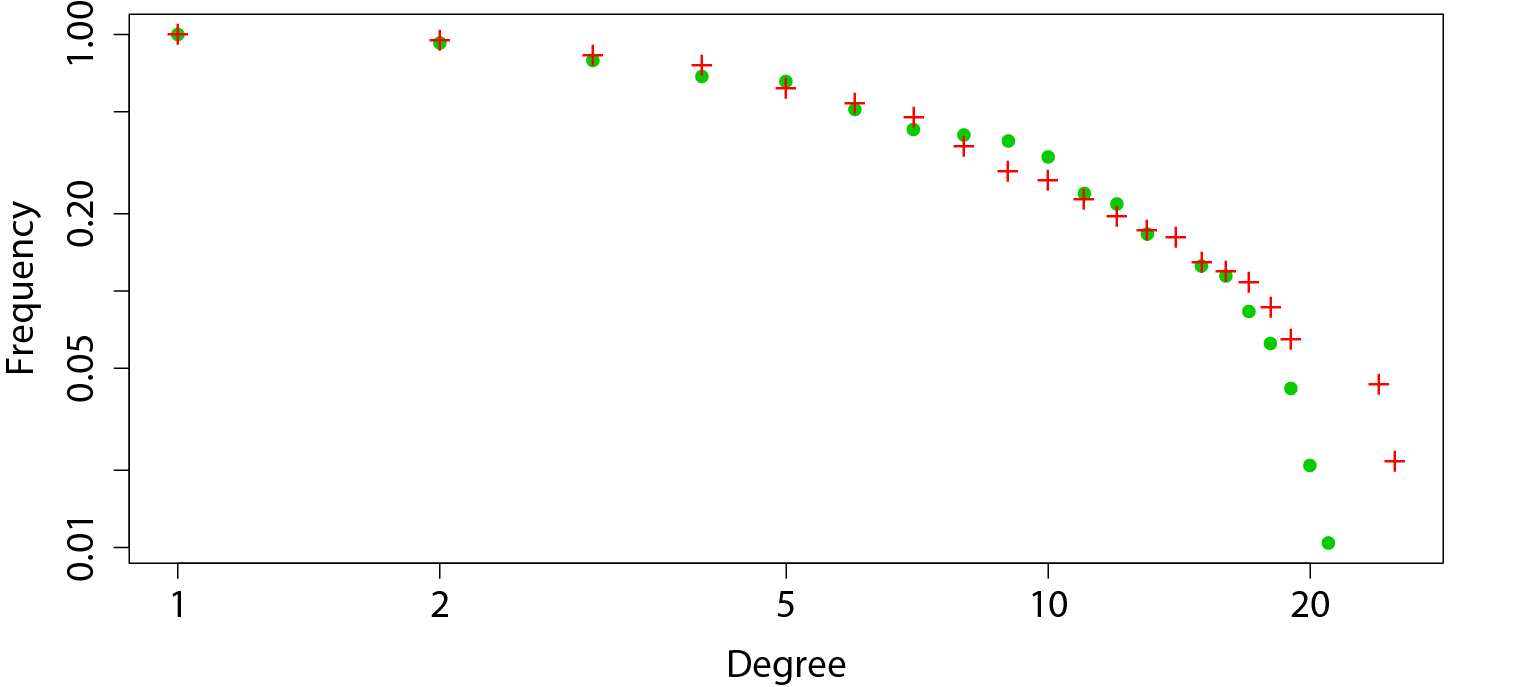}}\\
\subfloat[][]{%
\label{fig:bonf_dd-c}%
\includegraphics[width=0.5\textwidth]{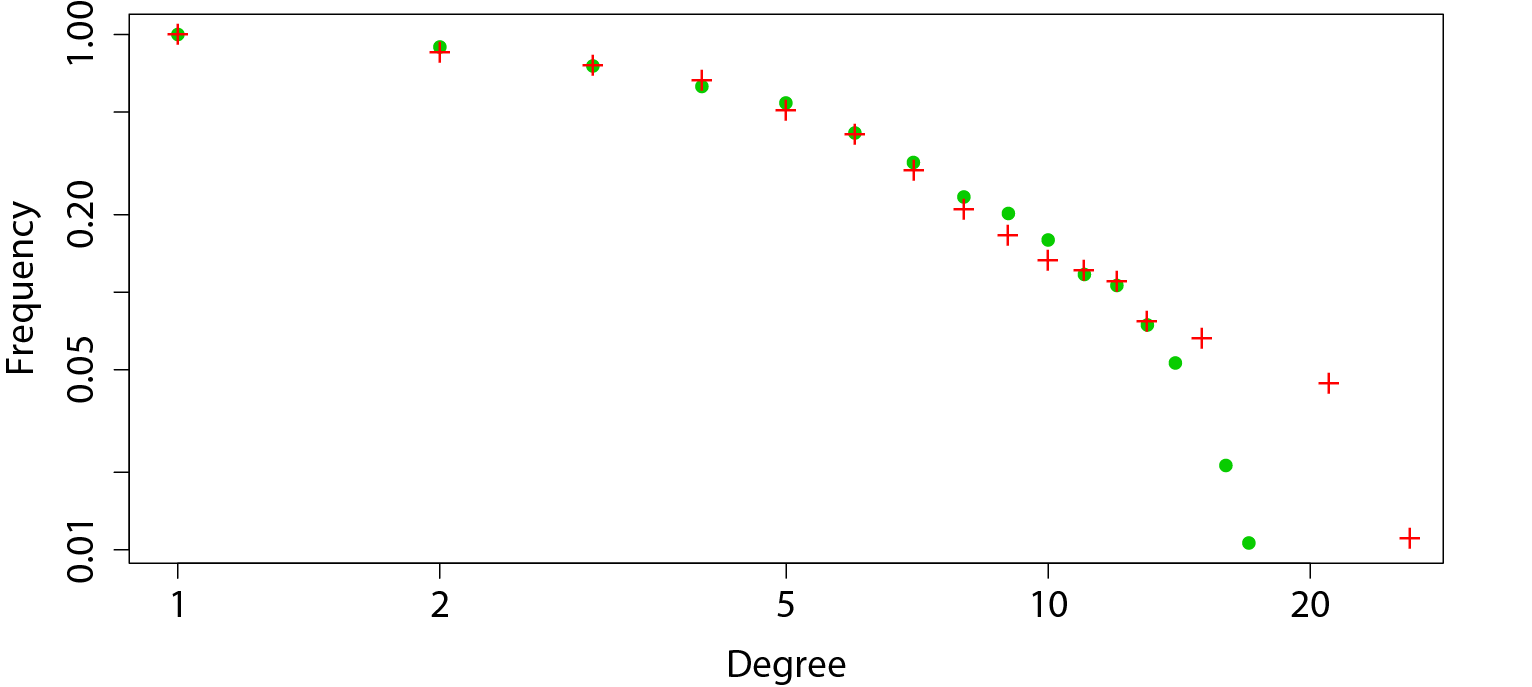}}%
\subfloat[][]{%
\label{fig:bonf_dd-d}%
\includegraphics[width=0.5\textwidth]{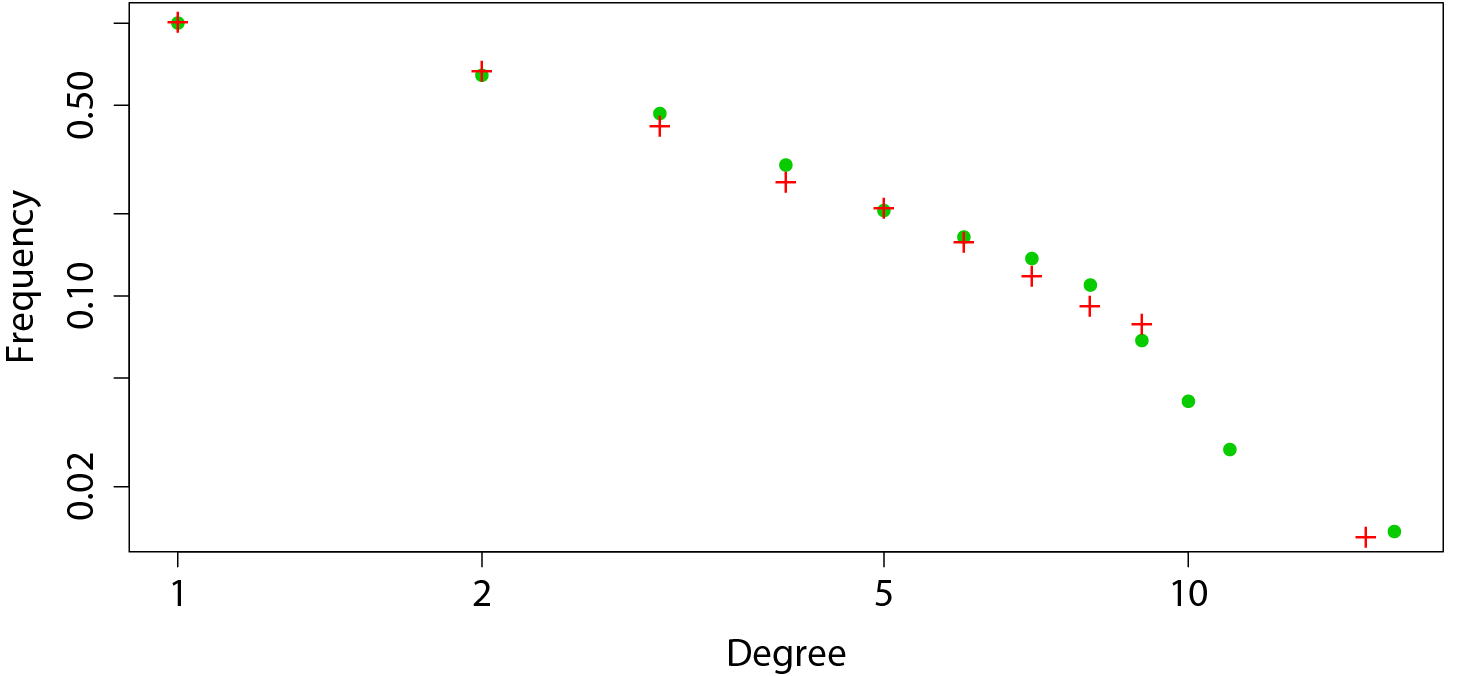}}\\
\subfloat[][]{%
\label{fig:bonf_dd-e}%
\includegraphics[width=0.5\textwidth]{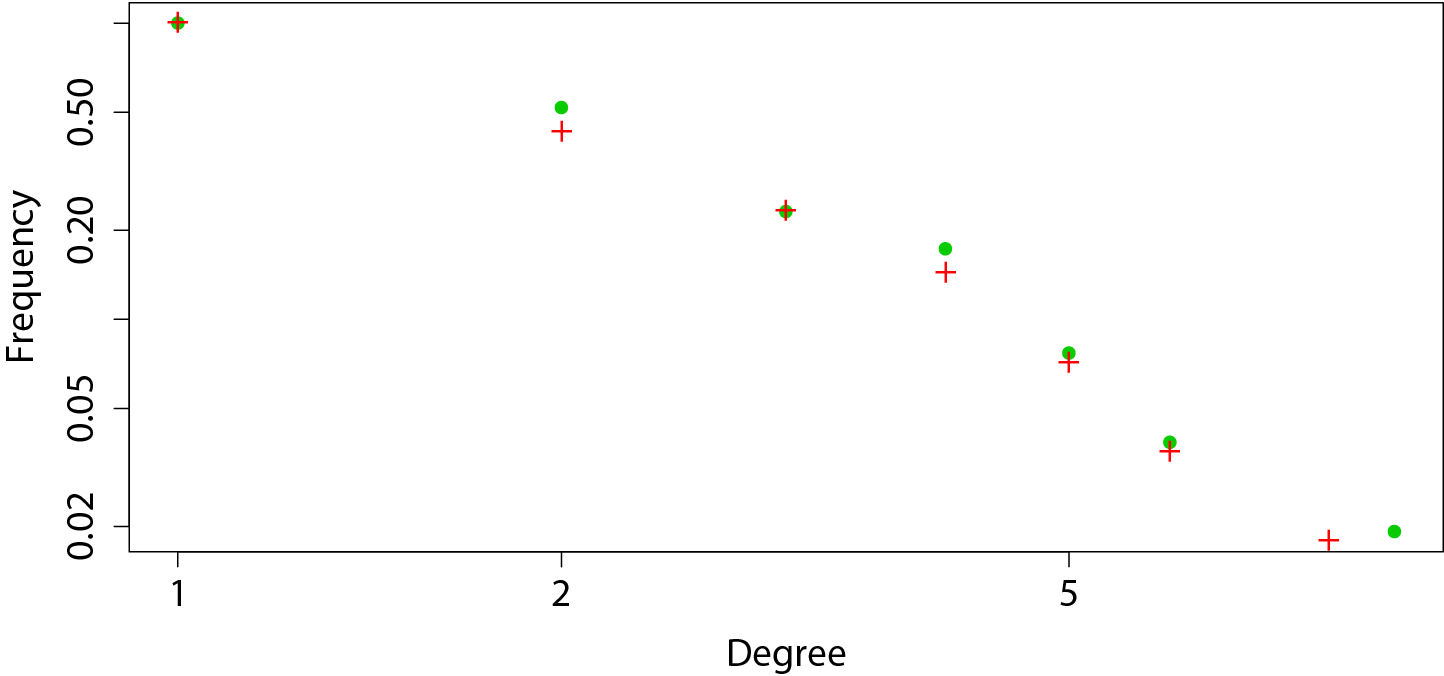}}%
\subfloat[][]{%
\label{fig:bonf_dd-f}%
\includegraphics[width=0.5\textwidth]{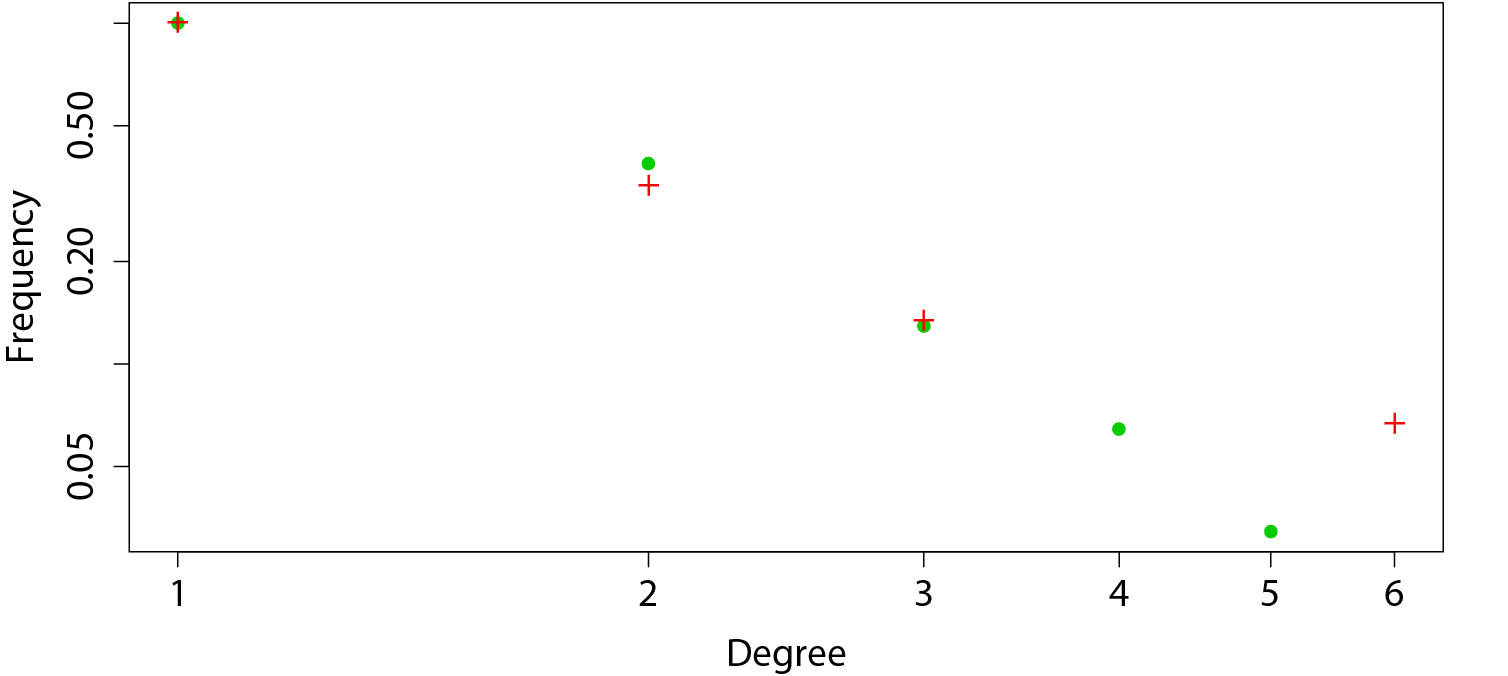}}%
\caption[Degree]{Degree distributions (in degree as dots, out degree as crosses) for Bonferroni networks based on significant transfer entropy between pairs of NYSE 100 stocks for time lag $\lambda$ of:
\subref{fig:bonf_dd-a} one minute;
\subref{fig:bonf_dd-b} five minutes;
\subref{fig:bonf_dd-c} ten minutes;
\subref{fig:bonf_dd-d} twenty minutes;
\subref{fig:bonf_dd-e} thirty minutes; and
\subref{fig:bonf_dd-f} forty minutes. The distributions for small values of $\lambda$ are close to random graphs, but move close to scale free networks for large values of $\lambda$.}%
\label{fig:bonf_dd}%
\end{figure*}

In Fig.~\ref{fig:bonf_dd} we present degree distribution for the networks presented in Fig.~\ref{fig:te_netw}. In and out degrees have been presented separately. The former as dots, and the latter as crosses. All distributions are presented on log-log scales.

\begin{figure*}%
\centering
\subfloat[][]{%
\label{fig:degree_dist-a}%
\includegraphics[width=0.5\textwidth]{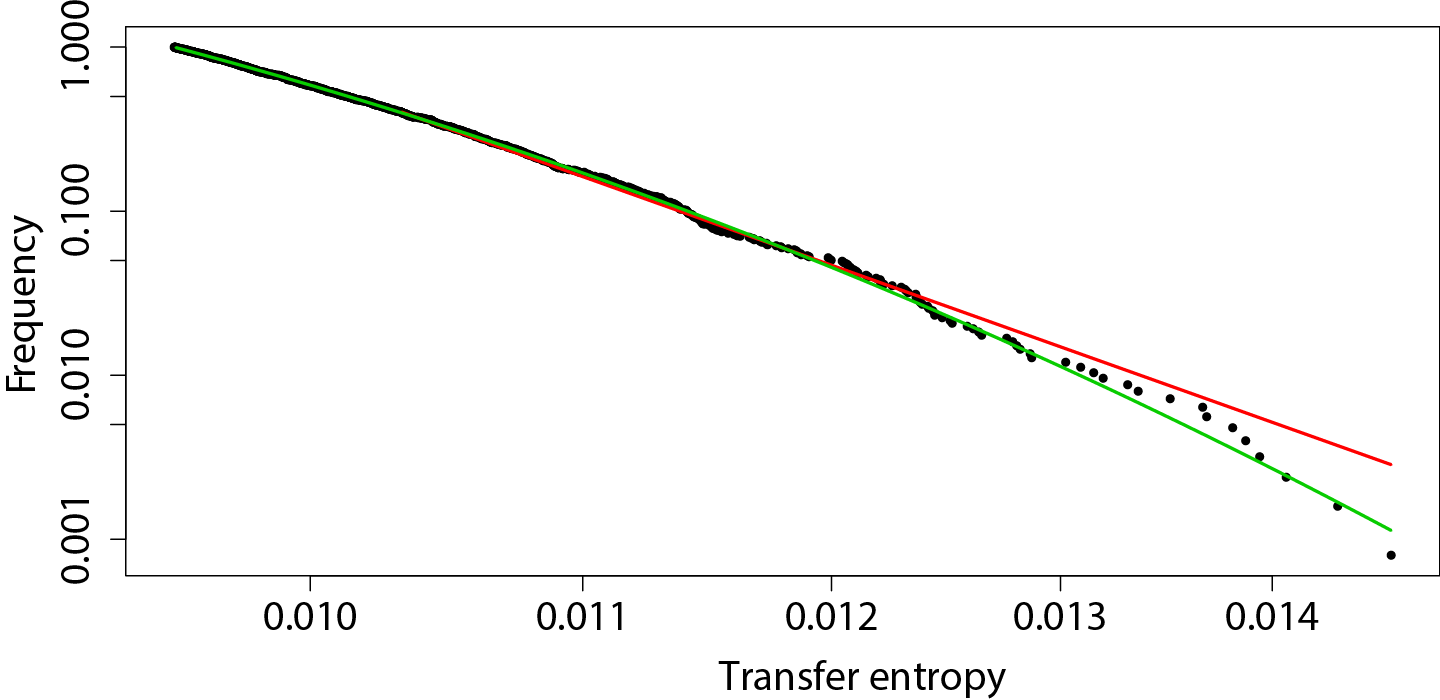}}%
\subfloat[][]{%
\label{fig:degree_dist-b}%
\includegraphics[width=0.5\textwidth]{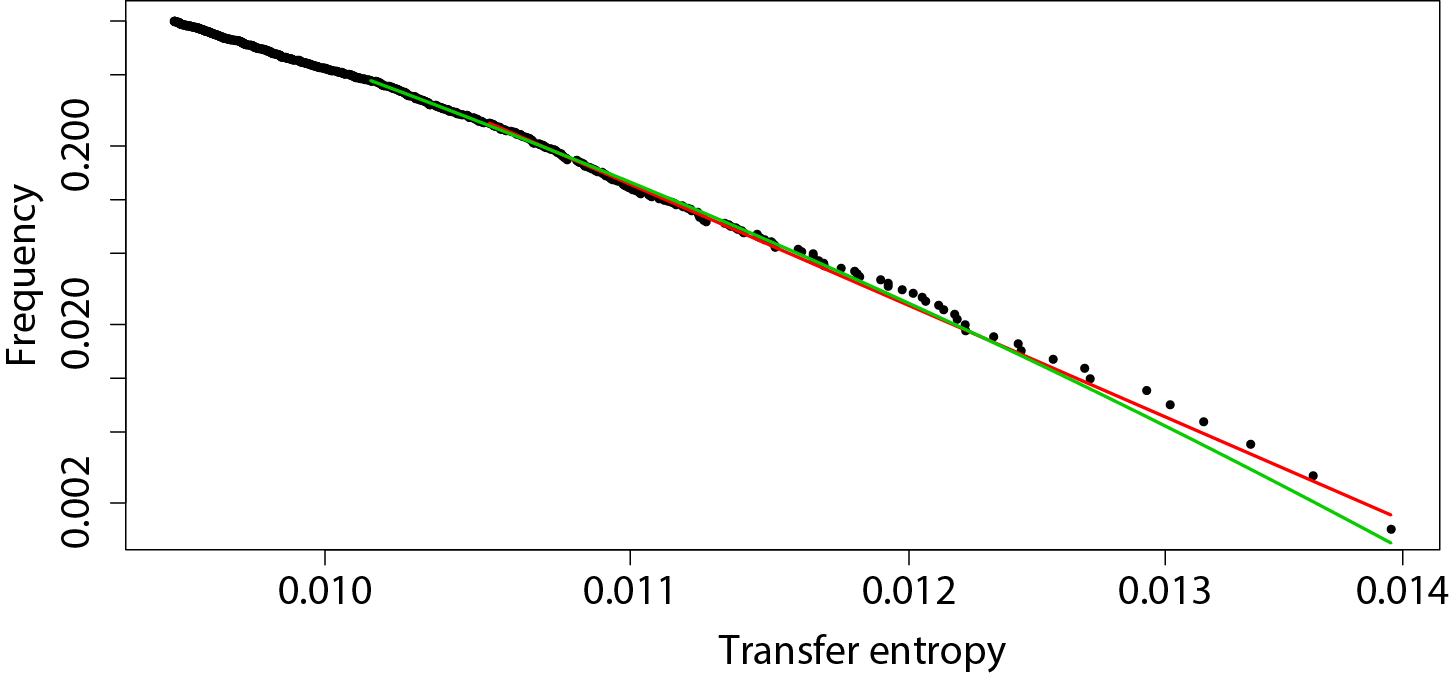}}\\
\subfloat[][]{%
\label{fig:degree_dist-c}%
\includegraphics[width=0.5\textwidth]{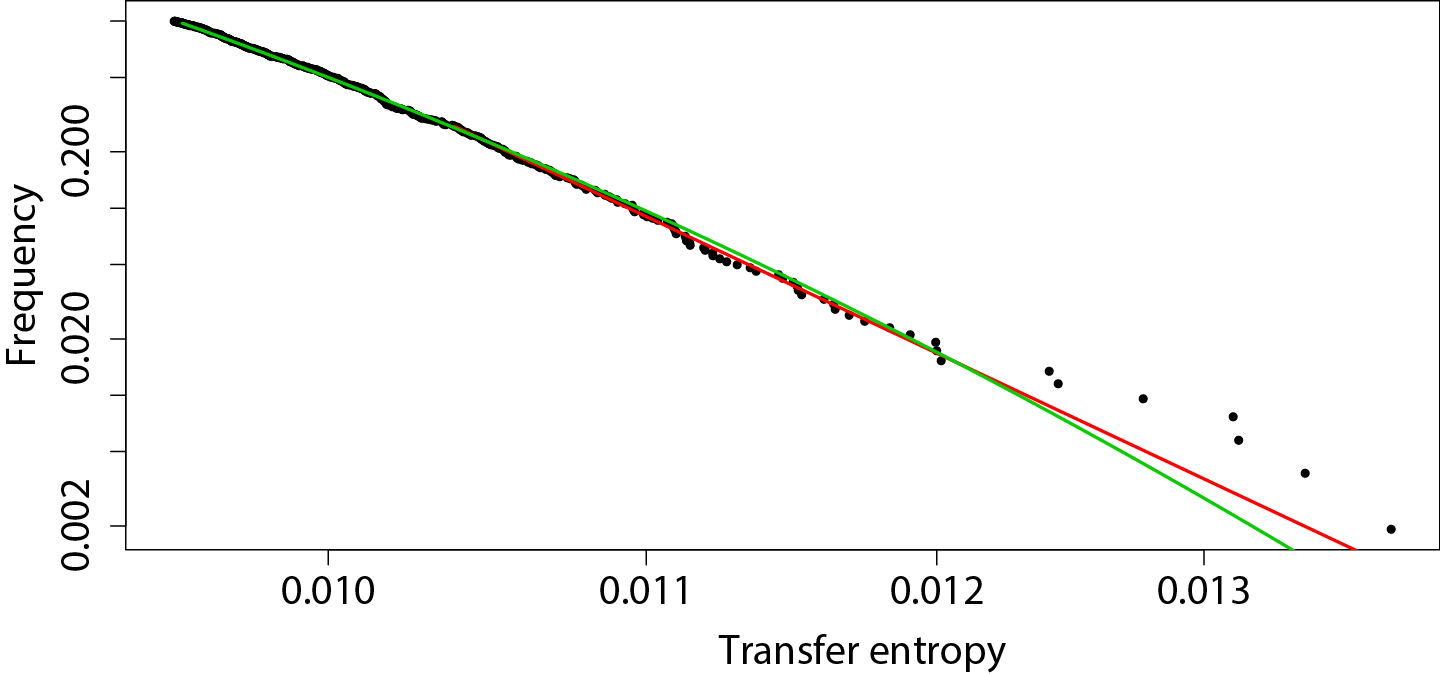}}%
\subfloat[][]{%
\label{fig:degree_dist-d}%
\includegraphics[width=0.5\textwidth]{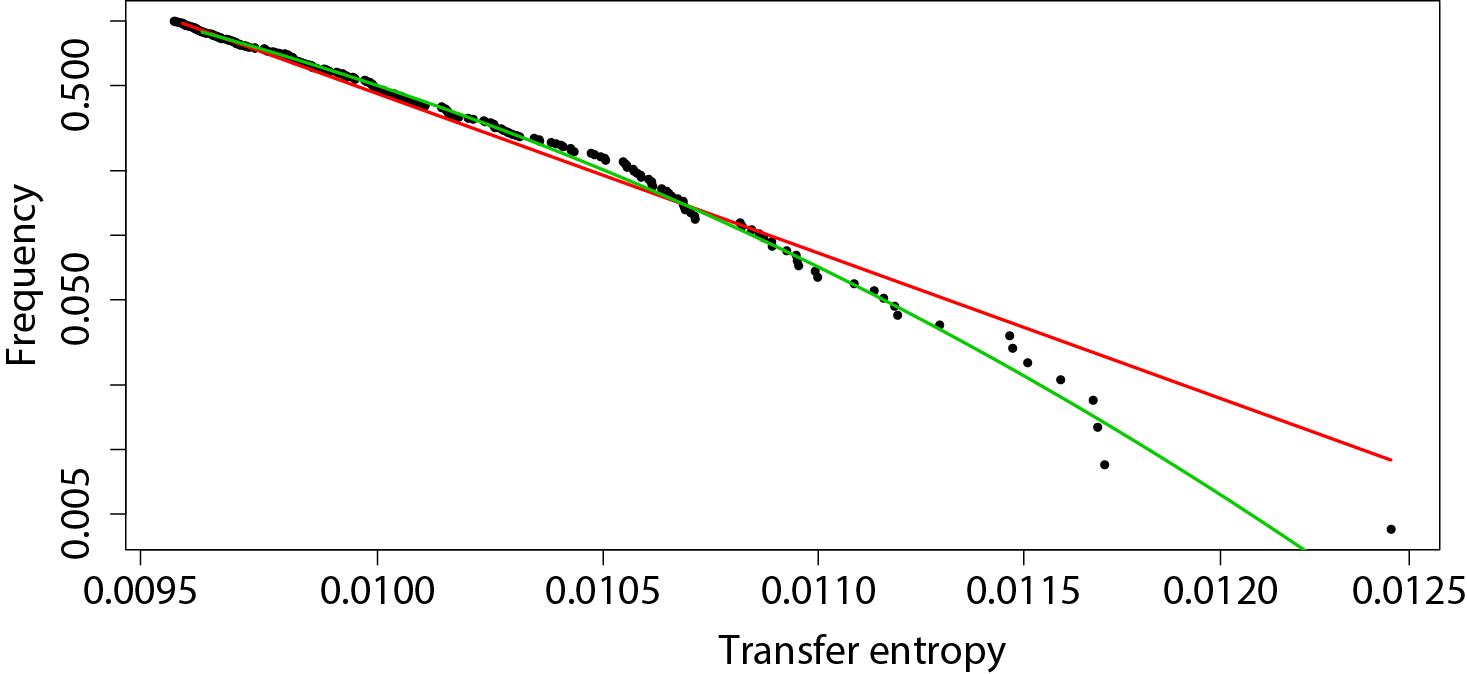}}\\
\subfloat[][]{%
\label{fig:degree_dist-e}%
\includegraphics[width=0.5\textwidth]{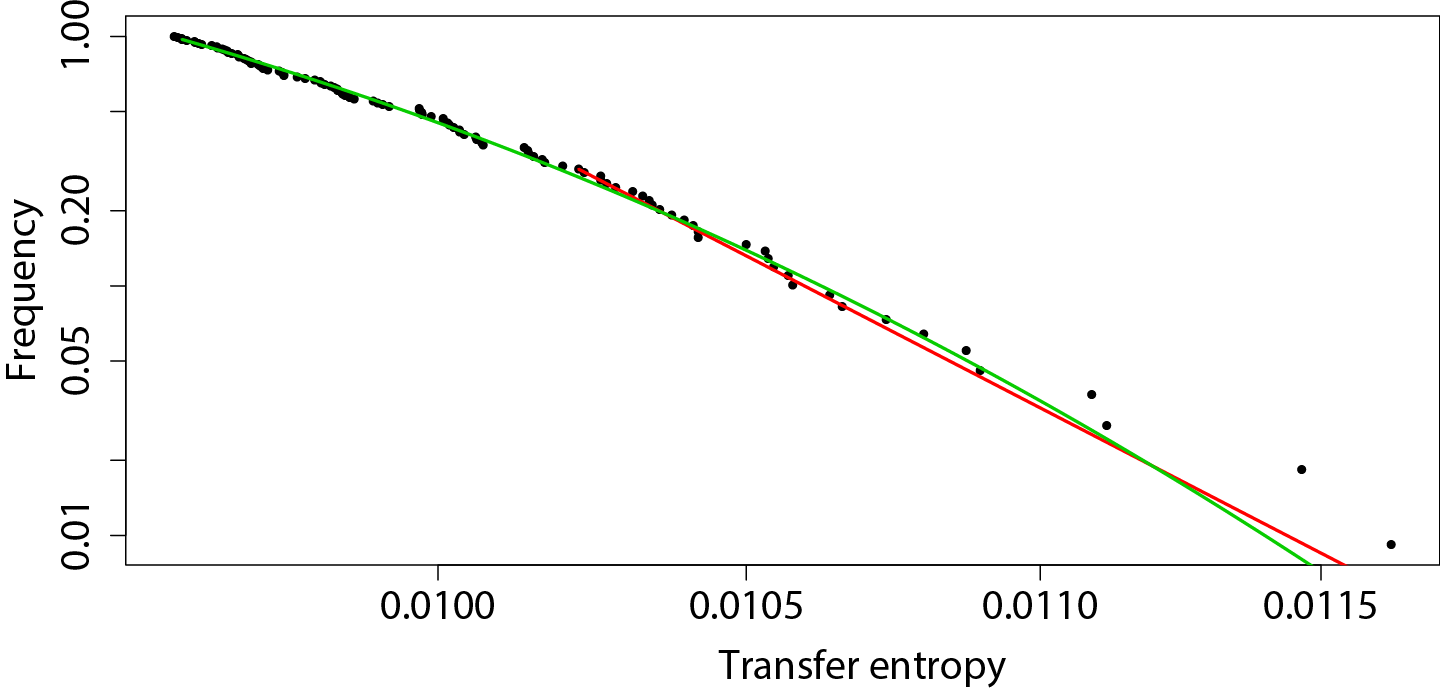}}%
\subfloat[][]{%
\label{fig:degree_dist-f}%
\includegraphics[width=0.5\textwidth]{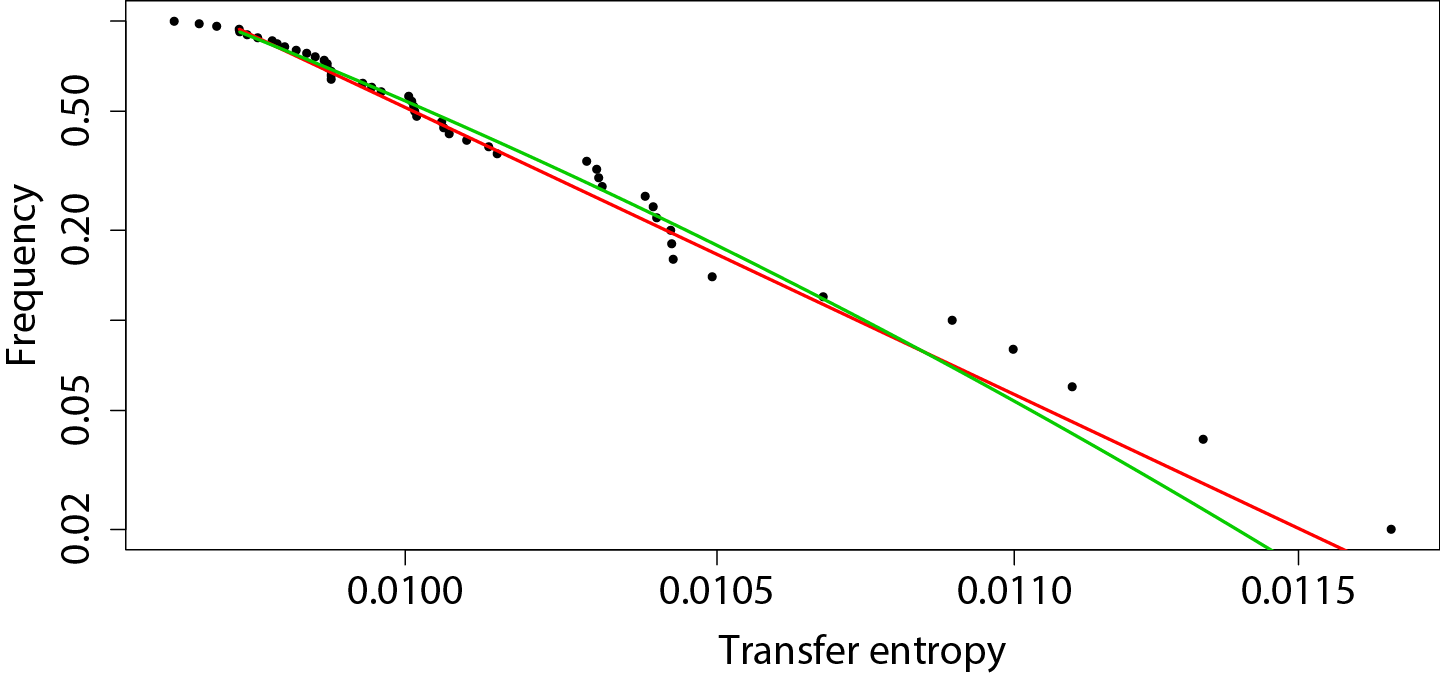}}%
\caption[Degree]{Degree distributions (with fitted power law and log-normal distributions) for statistically significant transfer entropy links between pairs of NYSE 100 stocks for time lag $\lambda$ of:
\subref{fig:degree_dist-a} one minute;
\subref{fig:degree_dist-b} five minutes;
\subref{fig:degree_dist-c} ten minutes;
\subref{fig:degree_dist-d} twenty minutes;
\subref{fig:degree_dist-e} thirty minutes; and
\subref{fig:degree_dist-f} forty minutes. There is only a handful of very strong transfer entropy relationships, the distributions are strongly fat-tailed.}%
\label{fig:degree_dist}%
\end{figure*}

In Fig.~\ref{fig:degree_dist} we present transfer entropy distributions for all statistically significant transfer entropy values within the studied set of stocks for time lags $\lambda$ of one, five, ten, twenty and forty minutes. These are presented on log-log scales. Additionally, these distributions have been fitted with appropriate power law and log-normal distributions.

\begin{figure}[tbh]
\centering
\includegraphics[width=0.4\textwidth]{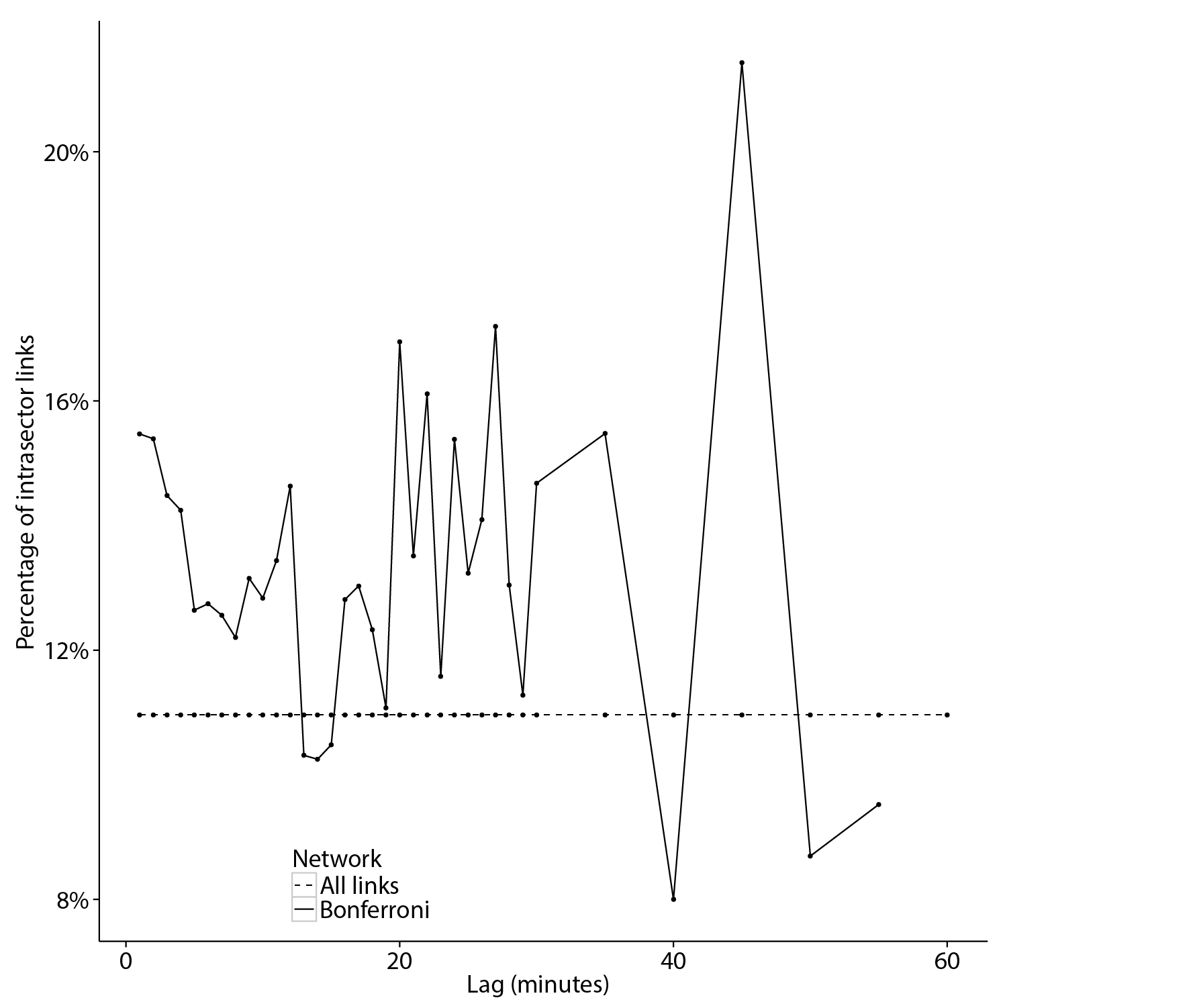}
\caption{Percentage of links between pairs of stocks belonging to the same economic sector in all links in Bonferroni and full unfiltered networks for various time lags $\lambda$. Bonferroni networks have only slightly increased percentage of intrasector links with regards to the unfiltered graph, which is in agreement with previous studies.}
\label{fig:intrasector}
\end{figure}

\begin{figure}[tbh]
\centering
\includegraphics[width=0.4\textwidth]{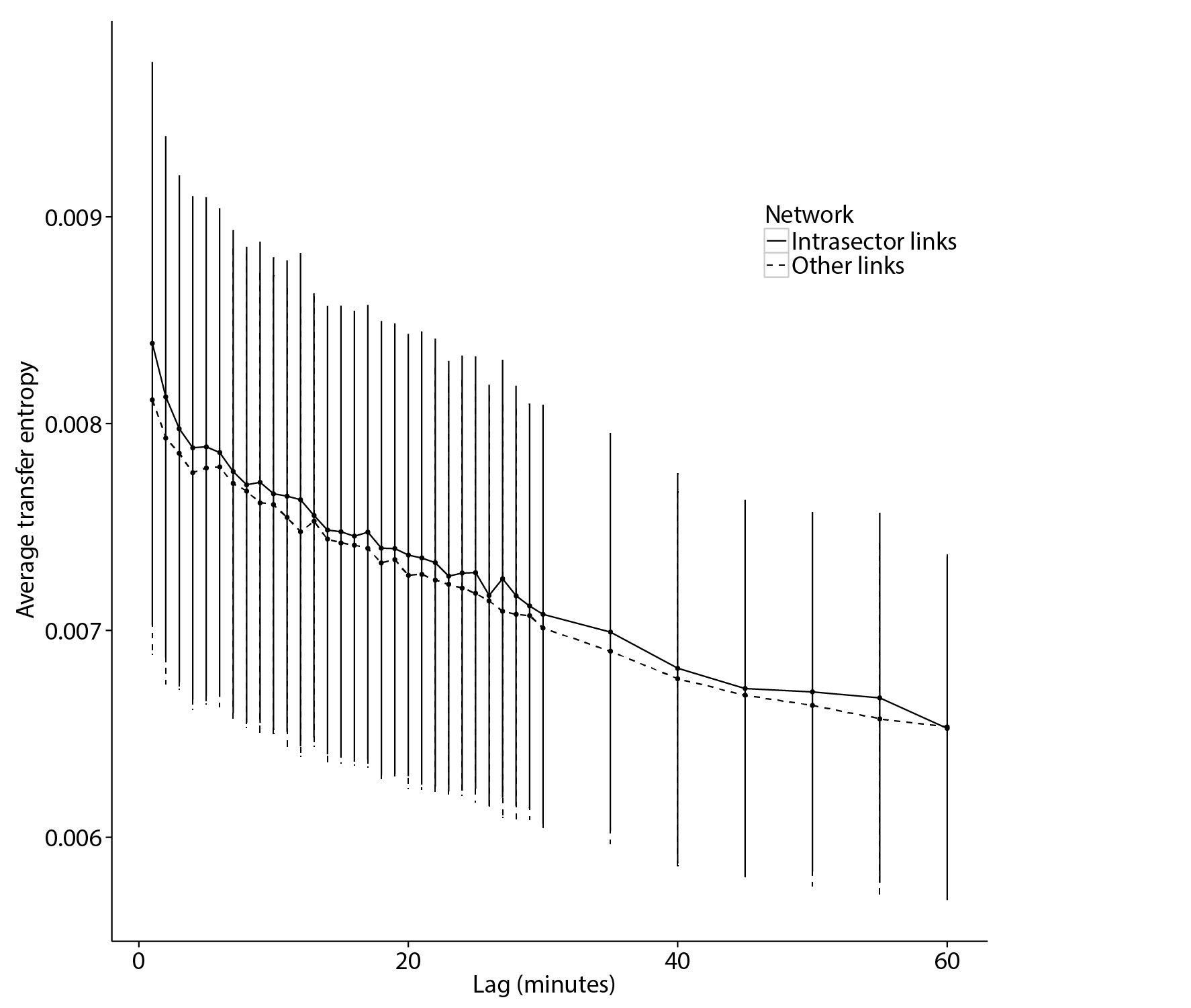}
\caption{Average magnitude of transfer entropy in unfiltered networks for all pairs of stocks, divided for the pairs between the same economic sector and other pairs, for various time lags $\lambda$. Error bars represent plus-or-minus one standard deviation. There is no statistical difference in transfer entropy between pairs of stocks belonging to the same sector and other pairs.}
\label{fig:instracomp}
\end{figure}

Further, in Fig.~\ref{fig:intrasector} we present the percentage of links between pairs of stocks belonging to the same economic sector in all links in Bonferroni and full unfiltered networks. This way we can observe whether the statistical filtering of significant transfer entropy--based relationships is based on the sector of economic activity of the studied stocks. For the same purpose in Fig.~\ref{fig:instracomp} we present the average transfer entropy for all pairs in the studied set divided into two categories: pairs of stocks belonging to the same economic sector and pairs of stocks belonging to two distinct economic sectors, for varying time lag $\lambda$. The error bars present plus-or-minus one standard deviation.

\section{Discussion}

We start the discussion by analysing the number of statistically significant links at different time lags ($\lambda$). If we analyse the interdependencies in a stock market at lag $\lambda=0$ (as is the case in most studies to date) then there are no arbitrage opportunities, as all price changes happen simultaneously. As such, there is no pressure to dissipate such synchronisation. And in fact we do observe very strong synchronisation within financial markets. The situation is different when analysing lead--lag effects however, that is synchronisation for time lags $\lambda>0$. There, statistically significant relationships constitute arbitrage opportunities, and as such may, at least in principle, be used for profit. In liquid markets we expect a pressure for these to be dissipated quickly, as market participants use these opportunities for profit. That is, we expect the number of statistically significant lead--lag links to decay very fast with increasing lag $\lambda$. In Fig.~\ref{fig:validated} we observe the number of statistically significant transfer entropy--based links within the studied market, for conservative Bonferroni corrections of $p$-value ($10^{-6}$). We see that the number of statistically significant links based on transfer entropy does dissipate rather quickly. Surprisingly, it does dissipate in a much slower manner as compared with our analysis using mutual information \cite{Fiedor:2014lag} however, and also slightly slower than in the analysis based on correlation \cite{Curme:2014}. Thus it appears that considering the effect of one stock on the changes in another leads to uncovering more relationships than simply considering the synchronisation between a pair of stocks, one of which is considered at a specific time lag. This may conceivably be due to the fact that the market participants find it easier to observe and use (in algorithmic trading) the opportunities emerging from the simpler relationships, which are captured by correlation or mutual information for lagged time series.

Here we present the effects of our validating the statistically significant transfer entropy values based on the $0.999999$ quantile of Gamma distribution with appropriate parameters. In Fig.~\ref{fig:tent} we can see the distribution of transfer entropy for all pairs of stocks within the studied set ($9506$) for time lags $\lambda$ of one, five, ten, twenty, thirty and forty minutes. The mentioned quantile (threshold), which denotes the point at which significant links begin, is presented as a vertical dashed line. The distribution approximates asymmetric Normal distribution, and is, as expected, moving to the left with increasing time lag $\lambda$. Thus we see, similarly to Fig.~\ref{fig:validated}, that with increasing time lag $\lambda$, the percentage of validated links is growing smaller. Additionally, in Fig.~\ref{fig:magnitude} we observe the average magnitude of the transfer entropy estimated for all links within two groups: all 9506 studied pairs of stocks (represented as the dashed line) and only the validated links constituting the Bonferroni networks (solid line). These are presented for all studied time lags $\lambda$. Additionally, error bars show the standard deviation (smaller for the significant links due to their cardinality). Again, we see that the average transfer entropy is getting smaller with increasing time lag for all studied links, but remains at a stable level for statistically significant links (due to stable Gamma quantile over all time lags). Here we note that the study of synchronisation based on Pearson's correlation of lagged time series \cite{Curme:2014} reported the links to be stable for the unfiltered networks. We find the results presented here to be closer to our expectations, as the relationships should be dissipating with increasing time lag, as explained above. What is more important, we see that the validated links are much stronger than the average transfer entropy for all studied pairs of stocks. Further, in Fig.~\ref{fig:correlation}, we see Pearson's correlation coefficients between transfer entropy for all pairs of studied stocks at different time lags $\lambda$. This is shown to analyse whether significant links persist at different time lags, or whether different causal relationships appear at different time lags. It appears the latter is the case, as the correlation coefficients are only weakly positive. This is in agreement with our intuition, as we would not expect the causality to hold for all time shifts.

Above we characterised transfer entropy for all pairs of stocks, or, in other words, for the unfiltered, full graphs. Here we turn to the analysis of the validated, statistically significant links between stocks, that is the filtered Bonferroni networks based on transfer entropy. The networks themselves for six time lags (one, five, ten, twenty, thirty and forty minutes) are presented on Fig.~\ref{fig:te_netw}. The same stock appears in the same place in all presented networks, and the colour of nodes is in line with the economic sector to which the stock belongs according to NYSE (12 sectors of economic activity). We can clearly observe the decay of links with increasing time lag when looking at the networks themselves. At this point we note that for small time lags the number of statistically significant links extends the number of links that standard topological methods would provide (e.g. Planar Maximally Filtered Graph). One may therefore be tempted to use such method, instead of applying threshold, to create more cohesive networks. There is a problem with such an approach however. We have created networks based on transfer entropy, which are roughly equivalent to Partial Correlation Planar Graphs introduced in Ref.~\cite{Kenett:2010} (these are not relevant in themselves, thus will not be presented). In the best case scenario we get a network where only $77.08\%$ of links are statistically significant. In most cases this number would drop to a much lower value. By adopting a topological method of creating filtered networks in this setting we would therefore be creating networks in which the amount of statistical noise would be very high. Thus we propose, as is proposed in other studies on similar topics, to use threshold for creating asynchronous financial networks.

The method of applying threshold does have a drawback however, in that the networks are not as cohesive as Minimally Spanning Trees or Planar Maximally Filtered Graphs. In other words the networks based on threshold are rarely small world networks, and rather often resemble random graphs. This in itself is not necessarily a bad thing, since we would not expect a causal structure to be the same as a static structure of financial markets. Nonetheless, in Fig.~\ref{fig:bonf_dd} we present degree distributions for Bonferroni networks at different time lags, as specified above. In degrees are presented with dots, and out degrees as crosses. All distributions are presented on log-log scales. We see that for small time lags $\lambda$ the networks approximate random graphs. In other words, for small lags the causal structure is not highly cohesive (the resulting distributions aren't well approximated by theoretical fat-tailed distributions such as power law or log-normal distributions). But we also observe that with increasing time lag the degree distributions move closer to fat-tailed distributions such as log-normal distribution or even power law. We postulate that this is due to the fact that for small time lags there is a lot of statistically significant links due to the synchronisation in the market, and these tend to be scattered around the market. The causal relationships which operate at higher time lags are usually strongly grounded in some economic basis, and thus create a more cohesive structure, similar to the standard synchronous scale free financial networks.

While we do not observe a scale free characteristic in the degree distributions of Bonferroni networks, we do observe a power law in the statistically significant transfer entropy values themselves. The distributions of these for the above-mentioned time lags are presented in Fig.~\ref{fig:degree_dist}, on log-log scales. We see that for all time lags the distribution is well approximated by either a power law or a log-normal distribution (both fitted and plotted). Thus we see that within the statistically significant causal relationship in New York's market only a handful are very strong, while most of them are only slightly above the threshold of statistical significance. This is desirable, as analyst may concentrate on the highly significant links, which they can be sure are not included in the networks due to badly chosen threshold. This also underscores that the choice of the threshold is not as important as it may appear at first glance, as any reasonable value will still retain the most important information within the constructed networks.

Finally, in Fig.~\ref{fig:intrasector} we present the percentage of pairs belonging to the same economic sector for all studied time lags $\lambda$ within two groups: all 9506 studied pairs (dashed line) and only the statistically significant pairs (solid line). We see that, unlike in the standard analysis of synchronous financial networks, such analysis does not lead to clusters defined by economic activity. The Bonferroni networks contain a higher percentage of intrasector links than the full graphs for most values of $\lambda$, but it is not an overwhelming difference in any case. This is in agreement with the results obtained with correlation--based methodology \cite{Curme:2014,Fiedor:2014lag}. Additionally, in Fig.~\ref{fig:instracomp} we show the average transfer entropy in both mentioned groups at all time lags $\lambda$, together with error bars denoting standard deviation. As can be seen, the difference is indeed insignificant, but it is always in favour of the intrasector links (solid lines). Such situation is not undesirable for two reasons. First, we would expect obvious links, which are due to economic sector, to be found by the analysts and dissipated immediately (before they can be included in such network). Second, we are more interested in finding the less obvious links between seemingly unconnected stocks. The presented analysis does not guarantee that all the links will be interesting and surprising, but it does make it plausible to find some such relationships using the presented methodology.

\section{Conclusions}

In this study we have enhanced the methodology for analysing asynchronous financial networks described in Refs.~\cite{Curme:2014,Fiedor:2014lag}. In particular, we have moved the analysis from synchronisation of lagged time series describing stock returns to the analysis of generalised (non-linear) causality (in the sense of Granger). We have presented a method of creating such networks based on the concept of transfer entropy, which naturally extends the methodology presented in our earlier study by exchanging mutual information to partial mutual information. We have applied this methodology to a dataset of NYSE 100 stocks, and presented the results. These seem to indicate that causal relationships are more prevalent on New York's market than lagged synchronisation relationships. We have observed, as expected, that the number of statistically significant links dissipates quickly (within an hour), and while the Bonferroni networks are not scale free (when studying the degrees of nodes, at least for small time lags), the distributions of significant transfer entropy values are indeed characterised best by fat-tailed distributions. We have also confirmed the earlier studies in that asynchronous links are rarely based on the sector of economic activity of the connected stocks. Further studies should look into enhancing this methodology, both in the method of quantifying causality and creating filtered networks, as well as applying it to other markets to show its usefulness and to present a comparative analysis of different markets in this respect.

\bibliography{prace}

\begin{thebibliography}{67}%
\makeatletter
\providecommand \@ifxundefined [1]{%
 \@ifx{#1\undefined}
}%
\providecommand \@ifnum [1]{%
 \ifnum #1\expandafter \@firstoftwo
 \else \expandafter \@secondoftwo
 \fi
}%
\providecommand \@ifx [1]{%
 \ifx #1\expandafter \@firstoftwo
 \else \expandafter \@secondoftwo
 \fi
}%
\providecommand \natexlab [1]{#1}%
\providecommand \enquote  [1]{``#1''}%
\providecommand \bibnamefont  [1]{#1}%
\providecommand \bibfnamefont [1]{#1}%
\providecommand \citenamefont [1]{#1}%
\providecommand \href@noop [0]{\@secondoftwo}%
\providecommand \href [0]{\begingroup \@sanitize@url \@href}%
\providecommand \@href[1]{\@@startlink{#1}\@@href}%
\providecommand \@@href[1]{\endgroup#1\@@endlink}%
\providecommand \@sanitize@url [0]{\catcode `\\12\catcode `\$12\catcode
  `\&12\catcode `\#12\catcode `\^12\catcode `\_12\catcode `\%12\relax}%
\providecommand \@@startlink[1]{}%
\providecommand \@@endlink[0]{}%
\providecommand \url  [0]{\begingroup\@sanitize@url \@url }%
\providecommand \@url [1]{\endgroup\@href {#1}{\urlprefix }}%
\providecommand \urlprefix  [0]{URL }%
\providecommand \Eprint [0]{\href }%
\providecommand \doibase [0]{http://dx.doi.org/}%
\providecommand \selectlanguage [0]{\@gobble}%
\providecommand \bibinfo  [0]{\@secondoftwo}%
\providecommand \bibfield  [0]{\@secondoftwo}%
\providecommand \translation [1]{[#1]}%
\providecommand \BibitemOpen [0]{}%
\providecommand \bibitemStop [0]{}%
\providecommand \bibitemNoStop [0]{.\EOS\space}%
\providecommand \EOS [0]{\spacefactor3000\relax}%
\providecommand \BibitemShut  [1]{\csname bibitem#1\endcsname}%
\let\auto@bib@innerbib\@empty
\bibitem [{\citenamefont {Mantegna}(1999)}]{Mantegna:1999}%
  \BibitemOpen
  \bibfield  {author} {\bibinfo {author} {\bibfnamefont {R.}~\bibnamefont
  {Mantegna}},\ }\href@noop {} {\bibfield  {journal} {\bibinfo  {journal}
  {{Eur. Phys. J. B}}\ }\textbf {\bibinfo {volume} {11}},\ \bibinfo {pages}
  {193} (\bibinfo {year} {1999})}\BibitemShut {NoStop}%
\bibitem [{\citenamefont {Samuelson}(1965)}]{Samuelson:1965}%
  \BibitemOpen
  \bibfield  {author} {\bibinfo {author} {\bibfnamefont {P.~A.}\ \bibnamefont
  {Samuelson}},\ }\href@noop {} {\bibfield  {journal} {\bibinfo  {journal}
  {{Ind. Manag. Rev.}}\ }\textbf {\bibinfo {volume} {6}},\ \bibinfo {pages}
  {41} (\bibinfo {year} {1965})}\BibitemShut {NoStop}%
\bibitem [{\citenamefont {Tobin}(1969)}]{Tobin:1969}%
  \BibitemOpen
  \bibfield  {author} {\bibinfo {author} {\bibfnamefont {J.}~\bibnamefont
  {Tobin}},\ }\href@noop {} {\bibfield  {journal} {\bibinfo  {journal} {{J.
  Money Credit Bank.}}\ }\textbf {\bibinfo {volume} {1}},\ \bibinfo {pages}
  {15} (\bibinfo {year} {1969})}\BibitemShut {NoStop}%
\bibitem [{\citenamefont {Lo}\ and\ \citenamefont {MacKinlay}(1988)}]{Lo:1988}%
  \BibitemOpen
  \bibfield  {author} {\bibinfo {author} {\bibfnamefont {A.}~\bibnamefont
  {Lo}}\ and\ \bibinfo {author} {\bibfnamefont {A.}~\bibnamefont {MacKinlay}},\
  }\href@noop {} {\bibfield  {journal} {\bibinfo  {journal} {{Rev. Financ.
  Stud.}}\ }\textbf {\bibinfo {volume} {1}},\ \bibinfo {pages} {41} (\bibinfo
  {year} {1988})}\BibitemShut {NoStop}%
\bibitem [{\citenamefont {Shmilovici}\ \emph {et~al.}(2003)\citenamefont
  {Shmilovici}, \citenamefont {Alon-Brimer},\ and\ \citenamefont
  {Hauser}}]{Shmilovici:2003}%
  \BibitemOpen
  \bibfield  {author} {\bibinfo {author} {\bibfnamefont {A.}~\bibnamefont
  {Shmilovici}}, \bibinfo {author} {\bibfnamefont {Y.}~\bibnamefont
  {Alon-Brimer}}, \ and\ \bibinfo {author} {\bibfnamefont {S.}~\bibnamefont
  {Hauser}},\ }\href@noop {} {\bibfield  {journal} {\bibinfo  {journal}
  {{Comput. Econ.}}\ }\textbf {\bibinfo {volume} {22}},\ \bibinfo {pages} {273}
  (\bibinfo {year} {2003})}\BibitemShut {NoStop}%
\bibitem [{\citenamefont {Fiedor}(2014{\natexlab{a}})}]{Fiedor:2014}%
  \BibitemOpen
  \bibfield  {author} {\bibinfo {author} {\bibfnamefont {P.}~\bibnamefont
  {Fiedor}},\ }in\ \href@noop {} {\emph {\bibinfo {booktitle} {{Proceedings of
  the IEEE Computational Intelligence for Financial Engineering \& Economics
  2014}}}},\ \bibinfo {editor} {edited by\ \bibinfo {editor} {\bibfnamefont
  {A.}~\bibnamefont {Serguieva}}, \bibinfo {editor} {\bibfnamefont
  {D.}~\bibnamefont {Maringer}}, \bibinfo {editor} {\bibfnamefont
  {V.}~\bibnamefont {Palade}}, \ and\ \bibinfo {editor} {\bibfnamefont {R.~J.}\
  \bibnamefont {Almeida}}}\ (\bibinfo  {publisher} {{IEEE}},\ \bibinfo
  {address} {London},\ \bibinfo {year} {2014})\ pp.\ \bibinfo {pages}
  {247--254}\BibitemShut {NoStop}%
\bibitem [{\citenamefont {Mandelbrot}(1963)}]{Mandelbrot:1963}%
  \BibitemOpen
  \bibfield  {author} {\bibinfo {author} {\bibfnamefont {B.~B.}\ \bibnamefont
  {Mandelbrot}},\ }\href@noop {} {\bibfield  {journal} {\bibinfo  {journal}
  {{J. Bus.}}\ }\textbf {\bibinfo {volume} {36}},\ \bibinfo {pages} {394}
  (\bibinfo {year} {1963})}\BibitemShut {NoStop}%
\bibitem [{\citenamefont {Kadanoff}(1971)}]{Kadanoff:1971}%
  \BibitemOpen
  \bibfield  {author} {\bibinfo {author} {\bibfnamefont {L.~P.}\ \bibnamefont
  {Kadanoff}},\ }\href@noop {} {\bibfield  {journal} {\bibinfo  {journal}
  {Simulation}\ }\textbf {\bibinfo {volume} {16}} (\bibinfo {year}
  {1971})}\BibitemShut {NoStop}%
\bibitem [{\citenamefont {Mantegna}(1991)}]{Mantegna:1991}%
  \BibitemOpen
  \bibfield  {author} {\bibinfo {author} {\bibfnamefont {R.~N.}\ \bibnamefont
  {Mantegna}},\ }\href@noop {} {\bibfield  {journal} {\bibinfo  {journal}
  {{Physica A}}\ }\textbf {\bibinfo {volume} {179}},\ \bibinfo {pages} {232}
  (\bibinfo {year} {1991})}\BibitemShut {NoStop}%
\bibitem [{\citenamefont {Cizeau}\ \emph {et~al.}(2001)\citenamefont {Cizeau},
  \citenamefont {Potters},\ and\ \citenamefont {Bouchaud}}]{Cizeau:2001}%
  \BibitemOpen
  \bibfield  {author} {\bibinfo {author} {\bibfnamefont {P.}~\bibnamefont
  {Cizeau}}, \bibinfo {author} {\bibfnamefont {M.}~\bibnamefont {Potters}}, \
  and\ \bibinfo {author} {\bibfnamefont {J.}~\bibnamefont {Bouchaud}},\
  }\href@noop {} {\bibfield  {journal} {\bibinfo  {journal} {{Quant. Financ.}}\
  }\textbf {\bibinfo {volume} {1}},\ \bibinfo {pages} {217} (\bibinfo {year}
  {2001})}\BibitemShut {NoStop}%
\bibitem [{\citenamefont {Forbes}\ and\ \citenamefont
  {Rigobon}(2002)}]{Forbes:2002}%
  \BibitemOpen
  \bibfield  {author} {\bibinfo {author} {\bibfnamefont {K.}~\bibnamefont
  {Forbes}}\ and\ \bibinfo {author} {\bibfnamefont {R.}~\bibnamefont
  {Rigobon}},\ }\href@noop {} {\bibfield  {journal} {\bibinfo  {journal} {{J.
  Financ.}}\ }\textbf {\bibinfo {volume} {57}},\ \bibinfo {pages} {2223}
  (\bibinfo {year} {2002})}\BibitemShut {NoStop}%
\bibitem [{\citenamefont {Podobnik}\ and\ \citenamefont
  {Stanley}(2008)}]{Podobnik:2008}%
  \BibitemOpen
  \bibfield  {author} {\bibinfo {author} {\bibfnamefont {B.}~\bibnamefont
  {Podobnik}}\ and\ \bibinfo {author} {\bibfnamefont {H.}~\bibnamefont
  {Stanley}},\ }\href@noop {} {\bibfield  {journal} {\bibinfo  {journal}
  {{Phys. Rev. Lett.}}\ }\textbf {\bibinfo {volume} {100}} (\bibinfo {year}
  {2008})}\BibitemShut {NoStop}%
\bibitem [{\citenamefont {Aste}\ \emph {et~al.}(2010)\citenamefont {Aste},
  \citenamefont {Shaw},\ and\ \citenamefont {Matteo}}]{Aste:2010}%
  \BibitemOpen
  \bibfield  {author} {\bibinfo {author} {\bibfnamefont {T.}~\bibnamefont
  {Aste}}, \bibinfo {author} {\bibfnamefont {W.}~\bibnamefont {Shaw}}, \ and\
  \bibinfo {author} {\bibfnamefont {T.~D.}\ \bibnamefont {Matteo}},\
  }\href@noop {} {\bibfield  {journal} {\bibinfo  {journal} {{New J. Phys.}}\
  }\textbf {\bibinfo {volume} {12}},\ \bibinfo {pages} {085009} (\bibinfo
  {year} {2010})}\BibitemShut {NoStop}%
\bibitem [{\citenamefont {Kenett}\ \emph {et~al.}(2012)\citenamefont {Kenett},
  \citenamefont {Preis}, \citenamefont {Gur-Gershgoren},\ and\ \citenamefont
  {Ben-Jacob}}]{Kenett:2012meta}%
  \BibitemOpen
  \bibfield  {author} {\bibinfo {author} {\bibfnamefont {D.}~\bibnamefont
  {Kenett}}, \bibinfo {author} {\bibfnamefont {T.}~\bibnamefont {Preis}},
  \bibinfo {author} {\bibfnamefont {G.}~\bibnamefont {Gur-Gershgoren}}, \ and\
  \bibinfo {author} {\bibfnamefont {E.}~\bibnamefont {Ben-Jacob}},\ }\href@noop
  {} {\bibfield  {journal} {\bibinfo  {journal} {{Europhys. Lett.}}\ }\textbf
  {\bibinfo {volume} {99}},\ \bibinfo {pages} {38001} (\bibinfo {year}
  {2012})}\BibitemShut {NoStop}%
\bibitem [{\citenamefont {Bonanno}\ \emph {et~al.}(2001)\citenamefont
  {Bonanno}, \citenamefont {Lillo},\ and\ \citenamefont
  {Mantegna}}]{Bonanno:2001}%
  \BibitemOpen
  \bibfield  {author} {\bibinfo {author} {\bibfnamefont {G.}~\bibnamefont
  {Bonanno}}, \bibinfo {author} {\bibfnamefont {F.}~\bibnamefont {Lillo}}, \
  and\ \bibinfo {author} {\bibfnamefont {R.}~\bibnamefont {Mantegna}},\
  }\href@noop {} {\bibfield  {journal} {\bibinfo  {journal} {{Quant. Financ.}}\
  }\textbf {\bibinfo {volume} {1}},\ \bibinfo {pages} {96} (\bibinfo {year}
  {2001})}\BibitemShut {NoStop}%
\bibitem [{\citenamefont {Tumminello}\ \emph
  {et~al.}(2007{\natexlab{a}})\citenamefont {Tumminello}, \citenamefont {Aste},
  \citenamefont {Matteo},\ and\ \citenamefont {Mantegna}}]{Tumminello:2007}%
  \BibitemOpen
  \bibfield  {author} {\bibinfo {author} {\bibfnamefont {M.}~\bibnamefont
  {Tumminello}}, \bibinfo {author} {\bibfnamefont {T.}~\bibnamefont {Aste}},
  \bibinfo {author} {\bibfnamefont {T.~D.}\ \bibnamefont {Matteo}}, \ and\
  \bibinfo {author} {\bibfnamefont {R.~N.}\ \bibnamefont {Mantegna}},\
  }\href@noop {} {\bibfield  {journal} {\bibinfo  {journal} {{Eur. Phys. J.
  B}}\ }\textbf {\bibinfo {volume} {55}},\ \bibinfo {pages} {209} (\bibinfo
  {year} {2007}{\natexlab{a}})}\BibitemShut {NoStop}%
\bibitem [{\citenamefont {Munnix}\ \emph {et~al.}(2010)\citenamefont {Munnix},
  \citenamefont {Schafer},\ and\ \citenamefont {Guhr}}]{Munnix:2010}%
  \BibitemOpen
  \bibfield  {author} {\bibinfo {author} {\bibfnamefont {M.}~\bibnamefont
  {Munnix}}, \bibinfo {author} {\bibfnamefont {R.}~\bibnamefont {Schafer}}, \
  and\ \bibinfo {author} {\bibfnamefont {T.}~\bibnamefont {Guhr}},\ }\href@noop
  {} {\bibfield  {journal} {\bibinfo  {journal} {{Physica A}}\ }\textbf
  {\bibinfo {volume} {389}},\ \bibinfo {pages} {4828} (\bibinfo {year}
  {2010})}\BibitemShut {NoStop}%
\bibitem [{\citenamefont {Billio}\ \emph {et~al.}(2012)\citenamefont {Billio},
  \citenamefont {Getmansky}, \citenamefont {Lo},\ and\ \citenamefont
  {Pelizzon}}]{Billio:2002}%
  \BibitemOpen
  \bibfield  {author} {\bibinfo {author} {\bibfnamefont {M.}~\bibnamefont
  {Billio}}, \bibinfo {author} {\bibfnamefont {M.}~\bibnamefont {Getmansky}},
  \bibinfo {author} {\bibfnamefont {A.}~\bibnamefont {Lo}}, \ and\ \bibinfo
  {author} {\bibfnamefont {L.}~\bibnamefont {Pelizzon}},\ }\href@noop {}
  {\bibfield  {journal} {\bibinfo  {journal} {{J. Financ. Econ.}}\ }\textbf
  {\bibinfo {volume} {104}},\ \bibinfo {pages} {535} (\bibinfo {year}
  {2012})}\BibitemShut {NoStop}%
\bibitem [{\citenamefont {Kenett}\ \emph {et~al.}(2010)\citenamefont {Kenett},
  \citenamefont {Tumminello}, \citenamefont {Madi}, \citenamefont
  {Gur-Gershgoren}, \citenamefont {Mantegna},\ and\ \citenamefont
  {Ben-Jacob}}]{Kenett:2010}%
  \BibitemOpen
  \bibfield  {author} {\bibinfo {author} {\bibfnamefont {D.}~\bibnamefont
  {Kenett}}, \bibinfo {author} {\bibfnamefont {M.}~\bibnamefont {Tumminello}},
  \bibinfo {author} {\bibfnamefont {A.}~\bibnamefont {Madi}}, \bibinfo {author}
  {\bibfnamefont {G.}~\bibnamefont {Gur-Gershgoren}}, \bibinfo {author}
  {\bibfnamefont {R.}~\bibnamefont {Mantegna}}, \ and\ \bibinfo {author}
  {\bibfnamefont {E.}~\bibnamefont {Ben-Jacob}},\ }\href@noop {} {\bibfield
  {journal} {\bibinfo  {journal} {{PloS one}}\ }\textbf {\bibinfo {volume}
  {5}},\ \bibinfo {pages} {e15032} (\bibinfo {year} {2010})}\BibitemShut
  {NoStop}%
\bibitem [{\citenamefont {Fiedor}(2014{\natexlab{b}})}]{Fiedor:2014a}%
  \BibitemOpen
  \bibfield  {author} {\bibinfo {author} {\bibfnamefont {P.}~\bibnamefont
  {Fiedor}},\ }\href@noop {} {\bibfield  {journal} {\bibinfo  {journal} {{Phys.
  Rev. E}}\ }\textbf {\bibinfo {volume} {89}},\ \bibinfo {pages} {052801}
  (\bibinfo {year} {2014}{\natexlab{b}})}\BibitemShut {NoStop}%
\bibitem [{\citenamefont {Laloux}\ \emph {et~al.}(2000)\citenamefont {Laloux},
  \citenamefont {Cizeau}, \citenamefont {Potters},\ and\ \citenamefont
  {Bouchaud}}]{Laloux:2000}%
  \BibitemOpen
  \bibfield  {author} {\bibinfo {author} {\bibfnamefont {L.}~\bibnamefont
  {Laloux}}, \bibinfo {author} {\bibfnamefont {P.}~\bibnamefont {Cizeau}},
  \bibinfo {author} {\bibfnamefont {M.}~\bibnamefont {Potters}}, \ and\
  \bibinfo {author} {\bibfnamefont {J.}~\bibnamefont {Bouchaud}},\ }\href@noop
  {} {\bibfield  {journal} {\bibinfo  {journal} {{Int. J. Theoretical Appl.
  Finance}}\ }\textbf {\bibinfo {volume} {3}},\ \bibinfo {pages} {391}
  (\bibinfo {year} {2000})}\BibitemShut {NoStop}%
\bibitem [{\citenamefont {Fenn}\ \emph {et~al.}(2011)\citenamefont {Fenn},
  \citenamefont {Porter}, \citenamefont {Williams}, \citenamefont {McDonald},
  \citenamefont {Johnson},\ and\ \citenamefont {Jones}}]{Fenn:2011}%
  \BibitemOpen
  \bibfield  {author} {\bibinfo {author} {\bibfnamefont {D.}~\bibnamefont
  {Fenn}}, \bibinfo {author} {\bibfnamefont {M.}~\bibnamefont {Porter}},
  \bibinfo {author} {\bibfnamefont {S.}~\bibnamefont {Williams}}, \bibinfo
  {author} {\bibfnamefont {M.}~\bibnamefont {McDonald}}, \bibinfo {author}
  {\bibfnamefont {N.}~\bibnamefont {Johnson}}, \ and\ \bibinfo {author}
  {\bibfnamefont {N.}~\bibnamefont {Jones}},\ }\href@noop {} {\bibfield
  {journal} {\bibinfo  {journal} {{Phys. Rev. E}}\ }\textbf {\bibinfo {volume}
  {84}},\ \bibinfo {pages} {026109} (\bibinfo {year} {2011})}\BibitemShut
  {NoStop}%
\bibitem [{\citenamefont {Bonanno}\ \emph {et~al.}(2003)\citenamefont
  {Bonanno}, \citenamefont {Caldarelli}, \citenamefont {Lillo},\ and\
  \citenamefont {Mantegna}}]{Bonanno:2003}%
  \BibitemOpen
  \bibfield  {author} {\bibinfo {author} {\bibfnamefont {G.}~\bibnamefont
  {Bonanno}}, \bibinfo {author} {\bibfnamefont {G.}~\bibnamefont {Caldarelli}},
  \bibinfo {author} {\bibfnamefont {F.}~\bibnamefont {Lillo}}, \ and\ \bibinfo
  {author} {\bibfnamefont {R.}~\bibnamefont {Mantegna}},\ }\href@noop {}
  {\bibfield  {journal} {\bibinfo  {journal} {{Phys. Rev. E}}\ }\textbf
  {\bibinfo {volume} {68}},\ \bibinfo {pages} {046130} (\bibinfo {year}
  {2003})}\BibitemShut {NoStop}%
\bibitem [{\citenamefont {Onnela}\ \emph {et~al.}(2003)\citenamefont {Onnela},
  \citenamefont {Chakraborti}, \citenamefont {Kaski},\ and\ \citenamefont
  {Kertesz}}]{Onnela:2003}%
  \BibitemOpen
  \bibfield  {author} {\bibinfo {author} {\bibfnamefont {J.-P.}\ \bibnamefont
  {Onnela}}, \bibinfo {author} {\bibfnamefont {A.}~\bibnamefont {Chakraborti}},
  \bibinfo {author} {\bibfnamefont {K.}~\bibnamefont {Kaski}}, \ and\ \bibinfo
  {author} {\bibfnamefont {J.}~\bibnamefont {Kertesz}},\ }\href@noop {}
  {\bibfield  {journal} {\bibinfo  {journal} {{Physica A}}\ }\textbf {\bibinfo
  {volume} {324}},\ \bibinfo {pages} {247} (\bibinfo {year}
  {2003})}\BibitemShut {NoStop}%
\bibitem [{\citenamefont {Tumminello}\ \emph {et~al.}(2005)\citenamefont
  {Tumminello}, \citenamefont {Aste}, \citenamefont {Matteo},\ and\
  \citenamefont {Mantegna}}]{Tumminello:2005}%
  \BibitemOpen
  \bibfield  {author} {\bibinfo {author} {\bibfnamefont {M.}~\bibnamefont
  {Tumminello}}, \bibinfo {author} {\bibfnamefont {T.}~\bibnamefont {Aste}},
  \bibinfo {author} {\bibfnamefont {T.~D.}\ \bibnamefont {Matteo}}, \ and\
  \bibinfo {author} {\bibfnamefont {R.~N.}\ \bibnamefont {Mantegna}},\
  }\href@noop {} {\bibfield  {journal} {\bibinfo  {journal} {{Proc. Natl. Acad.
  Sci. U.S.A.}}\ }\textbf {\bibinfo {volume} {102}},\ \bibinfo {pages} {10421}
  (\bibinfo {year} {2005})}\BibitemShut {NoStop}%
\bibitem [{\citenamefont {Tumminello}\ \emph
  {et~al.}(2007{\natexlab{b}})\citenamefont {Tumminello}, \citenamefont {Aste},
  \citenamefont {Matteo},\ and\ \citenamefont {Mantegna}}]{Tumminello:2010}%
  \BibitemOpen
  \bibfield  {author} {\bibinfo {author} {\bibfnamefont {M.}~\bibnamefont
  {Tumminello}}, \bibinfo {author} {\bibfnamefont {T.}~\bibnamefont {Aste}},
  \bibinfo {author} {\bibfnamefont {T.~D.}\ \bibnamefont {Matteo}}, \ and\
  \bibinfo {author} {\bibfnamefont {R.~N.}\ \bibnamefont {Mantegna}},\
  }\href@noop {} {\bibfield  {journal} {\bibinfo  {journal} {{Eur. Phys. J.
  B}}\ }\textbf {\bibinfo {volume} {55}},\ \bibinfo {pages} {209} (\bibinfo
  {year} {2007}{\natexlab{b}})}\BibitemShut {NoStop}%
\bibitem [{\citenamefont {Tumminello}\ \emph
  {et~al.}(2007{\natexlab{c}})\citenamefont {Tumminello}, \citenamefont
  {Coronnello}, \citenamefont {Lillo}, \citenamefont {Micciche},\ and\
  \citenamefont {Mantegna}}]{Coronnello:2007}%
  \BibitemOpen
  \bibfield  {author} {\bibinfo {author} {\bibfnamefont {M.}~\bibnamefont
  {Tumminello}}, \bibinfo {author} {\bibfnamefont {C.}~\bibnamefont
  {Coronnello}}, \bibinfo {author} {\bibfnamefont {F.}~\bibnamefont {Lillo}},
  \bibinfo {author} {\bibfnamefont {S.}~\bibnamefont {Micciche}}, \ and\
  \bibinfo {author} {\bibfnamefont {R.}~\bibnamefont {Mantegna}},\ }\href@noop
  {} {\bibfield  {journal} {\bibinfo  {journal} {{Int. J. Bifurcat. Chaos}}\
  }\textbf {\bibinfo {volume} {17}},\ \bibinfo {pages} {2319} (\bibinfo {year}
  {2007}{\natexlab{c}})}\BibitemShut {NoStop}%
\bibitem [{\citenamefont {Huth}\ and\ \citenamefont
  {Abergel}(2011)}]{Huth:2011}%
  \BibitemOpen
  \bibfield  {author} {\bibinfo {author} {\bibfnamefont {N.}~\bibnamefont
  {Huth}}\ and\ \bibinfo {author} {\bibfnamefont {F.}~\bibnamefont {Abergel}},\
  }\href@noop {} {\bibfield  {journal} {\bibinfo  {journal} {{ArXiV
  1111.7103}}\ } (\bibinfo {year} {2011})}\BibitemShut {NoStop}%
\bibitem [{\citenamefont {Curme}\ \emph {et~al.}(2014)\citenamefont {Curme},
  \citenamefont {Tumminello}, \citenamefont {Mantegna}, \citenamefont
  {Stanley},\ and\ \citenamefont {Kenett}}]{Curme:2014}%
  \BibitemOpen
  \bibfield  {author} {\bibinfo {author} {\bibfnamefont {C.}~\bibnamefont
  {Curme}}, \bibinfo {author} {\bibfnamefont {M.}~\bibnamefont {Tumminello}},
  \bibinfo {author} {\bibfnamefont {R.}~\bibnamefont {Mantegna}}, \bibinfo
  {author} {\bibfnamefont {H.}~\bibnamefont {Stanley}}, \ and\ \bibinfo
  {author} {\bibfnamefont {D.}~\bibnamefont {Kenett}},\ }\href@noop {}
  {\bibfield  {journal} {\bibinfo  {journal} {{ArXiV 1401.0462}}\ } (\bibinfo
  {year} {2014})}\BibitemShut {NoStop}%
\bibitem [{\citenamefont {Fiedor}(2014{\natexlab{c}})}]{Fiedor:2014lag}%
  \BibitemOpen
  \bibfield  {author} {\bibinfo {author} {\bibfnamefont {P.}~\bibnamefont
  {Fiedor}},\ }\href@noop {} {\bibfield  {journal} {\bibinfo  {journal} {{Eur.
  Phys. J. B}}\ }\textbf {\bibinfo {volume} {in print}} (\bibinfo {year}
  {2014}{\natexlab{c}})}\BibitemShut {NoStop}%
\bibitem [{\citenamefont {Brock}\ \emph {et~al.}(1991)\citenamefont {Brock},
  \citenamefont {Hsieh},\ and\ \citenamefont {LeBaron}}]{Brock:1991}%
  \BibitemOpen
  \bibfield  {author} {\bibinfo {author} {\bibfnamefont {W.~A.}\ \bibnamefont
  {Brock}}, \bibinfo {author} {\bibfnamefont {D.~A.}\ \bibnamefont {Hsieh}}, \
  and\ \bibinfo {author} {\bibfnamefont {B.}~\bibnamefont {LeBaron}},\
  }\href@noop {} {\emph {\bibinfo {title} {{Nonlinear Dynamics, Chaos, and
  Instability. Statistical Theory and Economic Evidence.}}}}\ (\bibinfo
  {publisher} {{MIT Press}},\ \bibinfo {address} {Cambridge},\ \bibinfo {year}
  {1991})\BibitemShut {NoStop}%
\bibitem [{\citenamefont {Qi}(1999)}]{Qi:1999}%
  \BibitemOpen
  \bibfield  {author} {\bibinfo {author} {\bibfnamefont {M.}~\bibnamefont
  {Qi}},\ }\href@noop {} {\bibfield  {journal} {\bibinfo  {journal} {{J. Bus.
  Econ. Stat.}}\ }\textbf {\bibinfo {volume} {17}},\ \bibinfo {pages} {419}
  (\bibinfo {year} {1999})}\BibitemShut {NoStop}%
\bibitem [{\citenamefont {McMillan}(2001)}]{McMillan:2001}%
  \BibitemOpen
  \bibfield  {author} {\bibinfo {author} {\bibfnamefont {D.}~\bibnamefont
  {McMillan}},\ }\href@noop {} {\bibfield  {journal} {\bibinfo  {journal}
  {{Int. Rev. Econ. Financ.}}\ }\textbf {\bibinfo {volume} {10}},\ \bibinfo
  {pages} {353} (\bibinfo {year} {2001})}\BibitemShut {NoStop}%
\bibitem [{\citenamefont {Sornette}\ and\ \citenamefont
  {Andersen}(2002)}]{Sornette:2002}%
  \BibitemOpen
  \bibfield  {author} {\bibinfo {author} {\bibfnamefont {D.}~\bibnamefont
  {Sornette}}\ and\ \bibinfo {author} {\bibfnamefont {J.}~\bibnamefont
  {Andersen}},\ }\href@noop {} {\bibfield  {journal} {\bibinfo  {journal}
  {{Int. J. Mod. Phys. C}}\ }\textbf {\bibinfo {volume} {13}},\ \bibinfo
  {pages} {171} (\bibinfo {year} {2002})}\BibitemShut {NoStop}%
\bibitem [{\citenamefont {Oh}\ and\ \citenamefont {Kim}(2002)}]{Kim:2002}%
  \BibitemOpen
  \bibfield  {author} {\bibinfo {author} {\bibfnamefont {K.}~\bibnamefont
  {Oh}}\ and\ \bibinfo {author} {\bibfnamefont {K.}~\bibnamefont {Kim}},\
  }\href@noop {} {\bibfield  {journal} {\bibinfo  {journal} {{Expert Syst.
  Appl.}}\ }\textbf {\bibinfo {volume} {22}},\ \bibinfo {pages} {249} (\bibinfo
  {year} {2002})}\BibitemShut {NoStop}%
\bibitem [{\citenamefont {Franses}\ and\ \citenamefont
  {Dijk}(1996)}]{Franses:1996}%
  \BibitemOpen
  \bibfield  {author} {\bibinfo {author} {\bibfnamefont {P.~H.}\ \bibnamefont
  {Franses}}\ and\ \bibinfo {author} {\bibfnamefont {D.~V.}\ \bibnamefont
  {Dijk}},\ }\href@noop {} {\bibfield  {journal} {\bibinfo  {journal} {{J.
  Forecasting}}\ }\textbf {\bibinfo {volume} {15}},\ \bibinfo {pages} {229}
  (\bibinfo {year} {1996})}\BibitemShut {NoStop}%
\bibitem [{\citenamefont {Abhyankar}\ \emph {et~al.}(1995)\citenamefont
  {Abhyankar}, \citenamefont {Copeland},\ and\ \citenamefont
  {Wong}}]{Wong:1995}%
  \BibitemOpen
  \bibfield  {author} {\bibinfo {author} {\bibfnamefont {A.}~\bibnamefont
  {Abhyankar}}, \bibinfo {author} {\bibfnamefont {L.}~\bibnamefont {Copeland}},
  \ and\ \bibinfo {author} {\bibfnamefont {W.}~\bibnamefont {Wong}},\
  }\href@noop {} {\bibfield  {journal} {\bibinfo  {journal} {{Econ. J.}}\
  }\textbf {\bibinfo {volume} {105}},\ \bibinfo {pages} {864} (\bibinfo {year}
  {1995})}\BibitemShut {NoStop}%
\bibitem [{\citenamefont {Chen}(1996)}]{Chen:1996}%
  \BibitemOpen
  \bibfield  {author} {\bibinfo {author} {\bibfnamefont {P.}~\bibnamefont
  {Chen}},\ }\href@noop {} {\bibfield  {journal} {\bibinfo  {journal} {{Stud.
  Nonlinear Dyn. E.}}\ }\textbf {\bibinfo {volume} {1}},\ \bibinfo {pages} {87}
  (\bibinfo {year} {1996})}\BibitemShut {NoStop}%
\bibitem [{\citenamefont {Abhyankar}\ \emph {et~al.}(1997)\citenamefont
  {Abhyankar}, \citenamefont {Copeland},\ and\ \citenamefont
  {Wong}}]{Wong:1997}%
  \BibitemOpen
  \bibfield  {author} {\bibinfo {author} {\bibfnamefont {A.}~\bibnamefont
  {Abhyankar}}, \bibinfo {author} {\bibfnamefont {L.}~\bibnamefont {Copeland}},
  \ and\ \bibinfo {author} {\bibfnamefont {W.}~\bibnamefont {Wong}},\
  }\href@noop {} {\bibfield  {journal} {\bibinfo  {journal} {{J. Bus. Econ.
  Stat.}}\ }\textbf {\bibinfo {volume} {15}},\ \bibinfo {pages} {1} (\bibinfo
  {year} {1997})}\BibitemShut {NoStop}%
\bibitem [{\citenamefont {Ammermann}\ and\ \citenamefont
  {Patterson}(2003)}]{Ammermann:2003}%
  \BibitemOpen
  \bibfield  {author} {\bibinfo {author} {\bibfnamefont {P.~A.}\ \bibnamefont
  {Ammermann}}\ and\ \bibinfo {author} {\bibfnamefont {D.~M.}\ \bibnamefont
  {Patterson}},\ }\href@noop {} {\bibfield  {journal} {\bibinfo  {journal}
  {{Pac. Bas. Financ. J.}}\ }\textbf {\bibinfo {volume} {11}},\ \bibinfo
  {pages} {175} (\bibinfo {year} {2003})}\BibitemShut {NoStop}%
\bibitem [{\citenamefont {Hsieh}(1989)}]{Hsieh:1989}%
  \BibitemOpen
  \bibfield  {author} {\bibinfo {author} {\bibfnamefont {D.}~\bibnamefont
  {Hsieh}},\ }\href@noop {} {\bibfield  {journal} {\bibinfo  {journal} {{J.
  Bus.}}\ }\textbf {\bibinfo {volume} {62}},\ \bibinfo {pages} {339} (\bibinfo
  {year} {1989})}\BibitemShut {NoStop}%
\bibitem [{\citenamefont {Meese}\ and\ \citenamefont {Rose}(1991)}]{Rose:1991}%
  \BibitemOpen
  \bibfield  {author} {\bibinfo {author} {\bibfnamefont {R.}~\bibnamefont
  {Meese}}\ and\ \bibinfo {author} {\bibfnamefont {A.}~\bibnamefont {Rose}},\
  }\href@noop {} {\bibfield  {journal} {\bibinfo  {journal} {{Rev. Econ.
  Stud.}}\ }\textbf {\bibinfo {volume} {58}},\ \bibinfo {pages} {603} (\bibinfo
  {year} {1991})}\BibitemShut {NoStop}%
\bibitem [{\citenamefont {Brooks}(1996)}]{Brooks:1996}%
  \BibitemOpen
  \bibfield  {author} {\bibinfo {author} {\bibfnamefont {C.}~\bibnamefont
  {Brooks}},\ }\href@noop {} {\bibfield  {journal} {\bibinfo  {journal} {{Appl.
  Financ. Econ.}}\ }\textbf {\bibinfo {volume} {6}},\ \bibinfo {pages} {307}
  (\bibinfo {year} {1996})}\BibitemShut {NoStop}%
\bibitem [{\citenamefont {Qi}\ and\ \citenamefont {Wu}(2003)}]{Wu:2003}%
  \BibitemOpen
  \bibfield  {author} {\bibinfo {author} {\bibfnamefont {M.}~\bibnamefont
  {Qi}}\ and\ \bibinfo {author} {\bibfnamefont {Y.}~\bibnamefont {Wu}},\
  }\href@noop {} {\bibfield  {journal} {\bibinfo  {journal} {{J. Empir.
  Financ.}}\ }\textbf {\bibinfo {volume} {10}},\ \bibinfo {pages} {623}
  (\bibinfo {year} {2003})}\BibitemShut {NoStop}%
\bibitem [{\citenamefont {Cover}\ and\ \citenamefont
  {Thomas}(1991)}]{Cover:1991}%
  \BibitemOpen
  \bibfield  {author} {\bibinfo {author} {\bibfnamefont {T.}~\bibnamefont
  {Cover}}\ and\ \bibinfo {author} {\bibfnamefont {J.}~\bibnamefont {Thomas}},\
  }\href@noop {} {\emph {\bibinfo {title} {Elements of Information Theory}}}\
  (\bibinfo  {publisher} {John Wiley \& Sons},\ \bibinfo {year}
  {1991})\BibitemShut {NoStop}%
\bibitem [{\citenamefont {Chicharro}\ and\ \citenamefont
  {Ledberg}(2012)}]{Chicharro580051}%
  \BibitemOpen
  \bibfield  {author} {\bibinfo {author} {\bibfnamefont {D.}~\bibnamefont
  {Chicharro}}\ and\ \bibinfo {author} {\bibfnamefont {A.}~\bibnamefont
  {Ledberg}},\ }\href@noop {} {\bibfield  {journal} {\bibinfo  {journal} {Phys.
  Rev. E}\ }\textbf {\bibinfo {volume} {86}} (\bibinfo {year}
  {2012})}\BibitemShut {NoStop}%
\bibitem [{\citenamefont {Marschinski}\ and\ \citenamefont
  {Matassini}(2001)}]{Marschinski:2001}%
  \BibitemOpen
  \bibfield  {author} {\bibinfo {author} {\bibfnamefont {R.}~\bibnamefont
  {Marschinski}}\ and\ \bibinfo {author} {\bibfnamefont {L.}~\bibnamefont
  {Matassini}},\ }\href {http://EconPapers.repec.org/RePEc:zbw:dbrrns:014}
  {\emph {\bibinfo {title} {Financial markets as a complex system: A short time
  scale perspective}}},\ \bibinfo {type} {Research Notes}\ \bibinfo {number}
  {01-4}\ (\bibinfo  {institution} {Deutsche Bank Research},\ \bibinfo {year}
  {2001})\BibitemShut {NoStop}%
\bibitem [{\citenamefont {Reddy}\ and\ \citenamefont
  {Sebastin}(2009)}]{Reddy:2009}%
  \BibitemOpen
  \bibfield  {author} {\bibinfo {author} {\bibfnamefont {Y.~V.}\ \bibnamefont
  {Reddy}}\ and\ \bibinfo {author} {\bibfnamefont {A.}~\bibnamefont
  {Sebastin}},\ }\href@noop {} {\bibfield  {journal} {\bibinfo  {journal} {{J.
  Alternative Invest.}}\ }\textbf {\bibinfo {volume} {11}},\ \bibinfo {pages}
  {85} (\bibinfo {year} {2009})}\BibitemShut {NoStop}%
\bibitem [{\citenamefont {Dimpfl}\ and\ \citenamefont
  {Peter}(2012)}]{Dimpfl2012Using}%
  \BibitemOpen
  \bibfield  {author} {\bibinfo {author} {\bibfnamefont {T.}~\bibnamefont
  {Dimpfl}}\ and\ \bibinfo {author} {\bibfnamefont {F.~J.}\ \bibnamefont
  {Peter}},\ }\href {http://hdl.handle.net/10419/79614} {\emph {\bibinfo
  {title} {Using transfer entropy to measure information flows between
  financial markets}}},\ \bibinfo {type} {SFB 649 Discussion Paper}\ \bibinfo
  {number} {2012-051}\ (\bibinfo  {institution} {Humboldt-Universitat zu
  Berlin},\ \bibinfo {address} {Berlin},\ \bibinfo {year} {2012})\BibitemShut
  {NoStop}%
\bibitem [{\citenamefont {Li}\ \emph {et~al.}(2013)\citenamefont {Li},
  \citenamefont {Liang}, \citenamefont {Zhu}, \citenamefont {Sun},\ and\
  \citenamefont {Wu}}]{Li:2013}%
  \BibitemOpen
  \bibfield  {author} {\bibinfo {author} {\bibfnamefont {J.}~\bibnamefont
  {Li}}, \bibinfo {author} {\bibfnamefont {C.}~\bibnamefont {Liang}}, \bibinfo
  {author} {\bibfnamefont {X.}~\bibnamefont {Zhu}}, \bibinfo {author}
  {\bibfnamefont {X.}~\bibnamefont {Sun}}, \ and\ \bibinfo {author}
  {\bibfnamefont {D.}~\bibnamefont {Wu}},\ }\href@noop {} {\bibfield  {journal}
  {\bibinfo  {journal} {{Entropy}}\ }\textbf {\bibinfo {volume} {15}},\
  \bibinfo {pages} {5549} (\bibinfo {year} {2013})}\BibitemShut {NoStop}%
\bibitem [{\citenamefont {Baek}\ \emph {et~al.}(2005)\citenamefont {Baek},
  \citenamefont {Jung}, \citenamefont {Kwon},\ and\ \citenamefont
  {Moon}}]{Baek:2005}%
  \BibitemOpen
  \bibfield  {author} {\bibinfo {author} {\bibfnamefont {S.~K.}\ \bibnamefont
  {Baek}}, \bibinfo {author} {\bibfnamefont {W.-S.}\ \bibnamefont {Jung}},
  \bibinfo {author} {\bibfnamefont {O.}~\bibnamefont {Kwon}}, \ and\ \bibinfo
  {author} {\bibfnamefont {H.-T.}\ \bibnamefont {Moon}},\ }\href@noop {}
  {\bibfield  {journal} {\bibinfo  {journal} {{ArXiV:physics/0509014}}\ }
  (\bibinfo {year} {2005})}\BibitemShut {NoStop}%
\bibitem [{\citenamefont {Kwon}\ and\ \citenamefont {Yang}(2008)}]{Kwon:2008}%
  \BibitemOpen
  \bibfield  {author} {\bibinfo {author} {\bibfnamefont {O.}~\bibnamefont
  {Kwon}}\ and\ \bibinfo {author} {\bibfnamefont {J.~S.}\ \bibnamefont
  {Yang}},\ }\href@noop {} {\bibfield  {journal} {\bibinfo  {journal} {EPL}\
  }\textbf {\bibinfo {volume} {82}},\ \bibinfo {pages} {68003} (\bibinfo {year}
  {2008})}\BibitemShut {NoStop}%
\bibitem [{\citenamefont {Junior}(2013)}]{Junior:2013}%
  \BibitemOpen
  \bibfield  {author} {\bibinfo {author} {\bibfnamefont {L.~S.}\ \bibnamefont
  {Junior}},\ }\href {http://EconPapers.repec.org/RePEc:ibm:ibmecp:wpe_324}
  {\emph {\bibinfo {title} {Structure and causality relations in a global
  network of financial companies}}},\ \bibinfo {type} {Insper Working Papers}\
  (\bibinfo  {institution} {Insper Working Paper, Insper Instituto de Ensino e
  Pesquisa},\ \bibinfo {year} {2013})\BibitemShut {NoStop}%
\bibitem [{\citenamefont {Frenzel}\ and\ \citenamefont
  {Pompe}(2007)}]{Frenzel:2007}%
  \BibitemOpen
  \bibfield  {author} {\bibinfo {author} {\bibfnamefont {S.}~\bibnamefont
  {Frenzel}}\ and\ \bibinfo {author} {\bibfnamefont {B.}~\bibnamefont
  {Pompe}},\ }\href@noop {} {\bibfield  {journal} {\bibinfo  {journal} {{Phys.
  Rev. Lett.}}\ }\textbf {\bibinfo {volume} {99}},\ \bibinfo {pages} {204101}
  (\bibinfo {year} {2007})}\BibitemShut {NoStop}%
\bibitem [{\citenamefont {Hlavackova-Schindler}\ \emph
  {et~al.}(2007)\citenamefont {Hlavackova-Schindler}, \citenamefont {Palus},
  \citenamefont {Vejmelka},\ and\ \citenamefont
  {Bhattacharya}}]{Schindler:2007}%
  \BibitemOpen
  \bibfield  {author} {\bibinfo {author} {\bibfnamefont {K.}~\bibnamefont
  {Hlavackova-Schindler}}, \bibinfo {author} {\bibfnamefont {M.}~\bibnamefont
  {Palus}}, \bibinfo {author} {\bibfnamefont {M.}~\bibnamefont {Vejmelka}}, \
  and\ \bibinfo {author} {\bibfnamefont {J.}~\bibnamefont {Bhattacharya}},\
  }\href@noop {} {\bibfield  {journal} {\bibinfo  {journal} {{Phys. Rep.}}\
  }\textbf {\bibinfo {volume} {441}},\ \bibinfo {pages} {1} (\bibinfo {year}
  {2007})}\BibitemShut {NoStop}%
\bibitem [{\citenamefont {Beirlant}\ \emph {et~al.}(1997)\citenamefont
  {Beirlant}, \citenamefont {Dudewicz}, \citenamefont {Gyorfi},\ and\
  \citenamefont {van~der Meulen}}]{Beirlant:1997}%
  \BibitemOpen
  \bibfield  {author} {\bibinfo {author} {\bibfnamefont {J.}~\bibnamefont
  {Beirlant}}, \bibinfo {author} {\bibfnamefont {E.}~\bibnamefont {Dudewicz}},
  \bibinfo {author} {\bibfnamefont {L.}~\bibnamefont {Gyorfi}}, \ and\ \bibinfo
  {author} {\bibfnamefont {E.}~\bibnamefont {van~der Meulen}},\ }\href@noop {}
  {\bibfield  {journal} {\bibinfo  {journal} {{Int. J. Math. Stat. Sci.}}\
  }\textbf {\bibinfo {volume} {6}},\ \bibinfo {pages} {17} (\bibinfo {year}
  {1997})}\BibitemShut {NoStop}%
\bibitem [{\citenamefont {Darbellay}\ and\ \citenamefont
  {Vajda}(1999)}]{Darbellay:1999}%
  \BibitemOpen
  \bibfield  {author} {\bibinfo {author} {\bibfnamefont {G.}~\bibnamefont
  {Darbellay}}\ and\ \bibinfo {author} {\bibfnamefont {I.}~\bibnamefont
  {Vajda}},\ }\href@noop {} {\bibfield  {journal} {\bibinfo  {journal} {{IEEE
  T. Inform. Theory}}\ }\textbf {\bibinfo {volume} {45}},\ \bibinfo {pages}
  {1315} (\bibinfo {year} {1999})}\BibitemShut {NoStop}%
\bibitem [{\citenamefont {Paninski}(2003)}]{Paninski:2003}%
  \BibitemOpen
  \bibfield  {author} {\bibinfo {author} {\bibfnamefont {L.}~\bibnamefont
  {Paninski}},\ }\href@noop {} {\bibfield  {journal} {\bibinfo  {journal}
  {{Neural Comput.}}\ }\textbf {\bibinfo {volume} {15}},\ \bibinfo {pages}
  {1191} (\bibinfo {year} {2003})}\BibitemShut {NoStop}%
\bibitem [{\citenamefont {Daub}\ \emph {et~al.}(2004)\citenamefont {Daub},
  \citenamefont {Steuer}, \citenamefont {Selbig},\ and\ \citenamefont
  {Kloska}}]{Daub:2004}%
  \BibitemOpen
  \bibfield  {author} {\bibinfo {author} {\bibfnamefont {C.}~\bibnamefont
  {Daub}}, \bibinfo {author} {\bibfnamefont {R.}~\bibnamefont {Steuer}},
  \bibinfo {author} {\bibfnamefont {J.}~\bibnamefont {Selbig}}, \ and\ \bibinfo
  {author} {\bibfnamefont {S.}~\bibnamefont {Kloska}},\ }\href@noop {}
  {\bibfield  {journal} {\bibinfo  {journal} {{BCM Bioinformatics}}\ }\textbf
  {\bibinfo {volume} {5}},\ \bibinfo {pages} {118} (\bibinfo {year}
  {2004})}\BibitemShut {NoStop}%
\bibitem [{\citenamefont {Nemenman}\ \emph {et~al.}(2004)\citenamefont
  {Nemenman}, \citenamefont {Bialek},\ and\ \citenamefont {de~Ruyter~van
  Steveninck}}]{Nemenman:2004}%
  \BibitemOpen
  \bibfield  {author} {\bibinfo {author} {\bibfnamefont {W.}~\bibnamefont
  {Nemenman}}, \bibinfo {author} {\bibfnamefont {W.}~\bibnamefont {Bialek}}, \
  and\ \bibinfo {author} {\bibfnamefont {R.}~\bibnamefont {de~Ruyter~van
  Steveninck}},\ }\href@noop {} {\bibfield  {journal} {\bibinfo  {journal}
  {{Phys. Rev. E}}\ }\textbf {\bibinfo {volume} {69}},\ \bibinfo {pages}
  {056111} (\bibinfo {year} {2004})}\BibitemShut {NoStop}%
\bibitem [{\citenamefont {Bonachela}\ \emph {et~al.}(2008)\citenamefont
  {Bonachela}, \citenamefont {H},\ and\ \citenamefont
  {Munoz}}]{Bonachela:2008}%
  \BibitemOpen
  \bibfield  {author} {\bibinfo {author} {\bibfnamefont {J.}~\bibnamefont
  {Bonachela}}, \bibinfo {author} {\bibfnamefont {H.}~\bibnamefont {H}}, \ and\
  \bibinfo {author} {\bibfnamefont {M.}~\bibnamefont {Munoz}},\ }\href@noop {}
  {\bibfield  {journal} {\bibinfo  {journal} {{J. Phys. A: Math. Theor.}}\
  }\textbf {\bibinfo {volume} {41}},\ \bibinfo {pages} {202001} (\bibinfo
  {year} {2008})}\BibitemShut {NoStop}%
\bibitem [{\citenamefont {Schurmann}\ and\ \citenamefont
  {Grassberger}(1996)}]{Schurmann:1996}%
  \BibitemOpen
  \bibfield  {author} {\bibinfo {author} {\bibfnamefont {T.}~\bibnamefont
  {Schurmann}}\ and\ \bibinfo {author} {\bibfnamefont {P.}~\bibnamefont
  {Grassberger}},\ }\href@noop {} {\bibfield  {journal} {\bibinfo  {journal}
  {{Chaos}}\ }\textbf {\bibinfo {volume} {6}},\ \bibinfo {pages} {414}
  (\bibinfo {year} {1996})}\BibitemShut {NoStop}%
\bibitem [{\citenamefont {Efron}\ and\ \citenamefont
  {Tibshirani}(1993)}]{Efron:1993}%
  \BibitemOpen
  \bibfield  {author} {\bibinfo {author} {\bibfnamefont {B.}~\bibnamefont
  {Efron}}\ and\ \bibinfo {author} {\bibfnamefont {R.}~\bibnamefont
  {Tibshirani}},\ }\href@noop {} {\emph {\bibinfo {title} {{An introduction to
  the bootstrap}}}}\ (\bibinfo  {publisher} {{CRC press}},\ \bibinfo {year}
  {1993})\BibitemShut {NoStop}%
\bibitem [{\citenamefont {Goebel}\ \emph {et~al.}(2005)\citenamefont {Goebel},
  \citenamefont {Dawy}, \citenamefont {Hagenauer},\ and\ \citenamefont
  {Mueller}}]{Goebel:2005}%
  \BibitemOpen
  \bibfield  {author} {\bibinfo {author} {\bibfnamefont {B.}~\bibnamefont
  {Goebel}}, \bibinfo {author} {\bibfnamefont {Z.}~\bibnamefont {Dawy}},
  \bibinfo {author} {\bibfnamefont {J.}~\bibnamefont {Hagenauer}}, \ and\
  \bibinfo {author} {\bibfnamefont {J.}~\bibnamefont {Mueller}},\ }in\
  \href@noop {} {\emph {\bibinfo {booktitle} {Proceedings of IEEE International
  Conference on Communications ICC 2005}}}\ (\bibinfo {year} {2005})\ pp.\
  \bibinfo {pages} {1102--1106}\BibitemShut {NoStop}%
\bibitem [{\citenamefont {Dawy}\ \emph {et~al.}(2006)\citenamefont {Dawy},
  \citenamefont {Goebel}, \citenamefont {Hagenauer}, \citenamefont {Andreoli},
  \citenamefont {Meitinger},\ and\ \citenamefont {Mueller}}]{Dawy:2006}%
  \BibitemOpen
  \bibfield  {author} {\bibinfo {author} {\bibfnamefont {Z.}~\bibnamefont
  {Dawy}}, \bibinfo {author} {\bibfnamefont {B.}~\bibnamefont {Goebel}},
  \bibinfo {author} {\bibfnamefont {J.}~\bibnamefont {Hagenauer}}, \bibinfo
  {author} {\bibfnamefont {C.}~\bibnamefont {Andreoli}}, \bibinfo {author}
  {\bibfnamefont {T.}~\bibnamefont {Meitinger}}, \ and\ \bibinfo {author}
  {\bibfnamefont {J.}~\bibnamefont {Mueller}},\ }\href@noop {} {\bibfield
  {journal} {\bibinfo  {journal} {{IEEE/ACM Trans. Comput. Biol. Bioinf.}}\
  }\textbf {\bibinfo {volume} {3}},\ \bibinfo {pages} {47} (\bibinfo {year}
  {2006})}\BibitemShut {NoStop}%
\bibitem [{\citenamefont {Steuer}\ \emph {et~al.}(2001)\citenamefont {Steuer},
  \citenamefont {Molgedey}, \citenamefont {Ebeling},\ and\ \citenamefont
  {Jiménez-Monta\~{n}o}}]{Steuer:2001}%
  \BibitemOpen
  \bibfield  {author} {\bibinfo {author} {\bibfnamefont {R.}~\bibnamefont
  {Steuer}}, \bibinfo {author} {\bibfnamefont {L.}~\bibnamefont {Molgedey}},
  \bibinfo {author} {\bibfnamefont {W.}~\bibnamefont {Ebeling}}, \ and\
  \bibinfo {author} {\bibfnamefont {M.}~\bibnamefont {Jiménez-Monta\~{n}o}},\
  }\href@noop {} {\bibfield  {journal} {\bibinfo  {journal} {Eur. Phys. J. B}\
  }\textbf {\bibinfo {volume} {19}},\ \bibinfo {pages} {265} (\bibinfo {year}
  {2001})}\BibitemShut {NoStop}%
\bibitem [{\citenamefont {Navet}\ and\ \citenamefont
  {Chen}(2008)}]{Navet:2008}%
  \BibitemOpen
  \bibfield  {author} {\bibinfo {author} {\bibfnamefont {N.}~\bibnamefont
  {Navet}}\ and\ \bibinfo {author} {\bibfnamefont {S.-H.}\ \bibnamefont
  {Chen}},\ }in\ \href@noop {} {\emph {\bibinfo {booktitle} {{Natural Computing
  in Computational Finance}}}},\ \bibinfo {series} {Studies in Computational
  Intelligence}, Vol.\ \bibinfo {volume} {100},\ \bibinfo {editor} {edited by\
  \bibinfo {editor} {\bibfnamefont {T.}~\bibnamefont {Brabazon}}\ and\ \bibinfo
  {editor} {\bibfnamefont {M.}~\bibnamefont {O'Neill}}}\ (\bibinfo  {publisher}
  {Springer},\ \bibinfo {address} {Berlin Heidelberg},\ \bibinfo {year}
  {2008})\BibitemShut {NoStop}%
\end{thebibliography}%
\end{document}